\documentclass[%
 reprint,
 amsmath,amssymb,
 aps,
]{revtex4-2}

\usepackage{graphicx}
\usepackage{dcolumn}
\usepackage{bm}
\usepackage{caption}
\usepackage{array}
\usepackage{float}
\newcolumntype{P}[1]{>{\centering\arraybackslash}p{#1}}

\newcommand{\dy}[1]{\underline{\underline{\bm{#1}}}}

\newcommand{\RomanNumeralCaps}[1]{\MakeUppercase{\romannumeral #1}}
\begin{document}

\preprint{APS/123-QED}

\title{On the surface plasmonic waves excited by a dipole above anisotropic and spatially dispersive two-dimensional surfaces of infinite extent in planarly layered media}

\author{Minyu Gu}
 \email{guminyu@tamu.edu}
 \affiliation{%
 	Department of Electrical and Computer Engineering, Texas A\&M University, College Station, TX, 77843 USA.}
\author{Krzysztof A. Michalski}%
 \email{k-michalski@tamu.edu}
\affiliation{%
Department of Electrical and Computer Engineering, Texas A\&M University, College Station, TX, 77843 USA.}
%




\date{\today}

\begin{abstract}
The surface plasmonic waves excited by a vertical or horizontal oriented Hertzian dipole above anisotropic and spatially dispersive two-dimensional surfaces of infinite extent embedded in planarly layered uniaxial media is investigated using the dyadic Green function approach. The spectral-domain transmission line analogy Green function formulation and iso-frequency contours equations are derived. The methods to accurately and efficently evaluate the two-dimensional Fourier integral arisen from the spatial-domain Green function computation are also developed. To resolve the numerical inefficiency due to the highly oscillatory integrand and singularities of surface plasmonic waves possessing large wavenumber, two numerical strategies, the extrapolation of the real-axis integration combined with singularity subtraction, and the deformed vertical integration path, are proposed and applicable to a wide range of observation distance. As a demonstration of the proposed formulation, we compute the scattered fields of a vertical dipole above the graphene biased by drift current which exhibits significant spatial dispersion and show that its light-matter interaction can be significantly reinforced when placed above uniaxially epsilon-near-zero substrates. The proposed formulation may provide methodology for the computational analysis of two-dimensional materials and surface plasmonic waves.
\end{abstract}

\maketitle


\section{\label{sec:level1}Introduction}
Recent development in the photonics and THz electromagnetic waves arouses great interest in surface plasmonic waves (SPWs) propagating on innovative two-dimensional surfaces. It has been demonstrated that the intrinsic plasmons frequency of graphene with notable charge carrier concentration lies in the low THz frequency regime, and promises a significant breakthrough of unconventional devices \cite{ju2011graphene}. Various approaches have been taken to engineer surface plasmonic polarizations exits on the two-dimensional surfaces, Ref.~\cite{tamagnone2012reconfigurable} proposed a frequency-tuned graphene antenna to achieve high miniaturization and direct matching over a large frequency range. Ref.~\cite{gomez2015hyperbolic} deomstrates that densely-packed graphene ribbon array can produce a hyperbolic response which arises high directivity and extreme confinement of the supported SPWs. Ref.~\cite{morgado2018drift} reports that the graphene biased by drift current can break electromagnetic reciprocity and induce unidirectional SPWs. Ref.~\cite{dai2015graphene} illustrates that hyperbolic plasmon–phonon polaritons induced by graphene hybridized with hexagonal boron nitride can mitigate ohmic losses and surports longer SPWs propagating distance.
\par
Many works have demonstrated unconventional electromagnetic phenomena can exit on hybridizing the two-dimensional materials with the anisotropic bulk materials. However, there hasn't been a report on the perspective of the efficent numerical computation of dipole sources incident on such structures. The purpose of this work is to develop a general formulation and computational method to solve the surface plasmonic waves excited by Hertzian dipoles above anisotropic spatially dispersive two-dimensional surfaces in planarly layered uniaxial media using the dyadic Green function approach, which provides an accurate and efficient computational modeling of electromagnetic SPWs propagating on two-dimensional materials and metasurfaces. We start with representing the spectral-domain Maxwell equations with a transmission line analog equation and modeling the anisotropic conductive surfaces by an interface transform results in coupled Transverse Electric (TE) Transverse Magnetic (TM) modes. Next the methods to solve the spectral-domain Green function and the pole-free formulation of the iso-frequency contours (IFCs) are shown.
\par
To obtain the spatial-domain Green functions, two efficient evaluation strategies of conducting the two-dimensional Fourier integral in the cylindrical coordinate system are posed. The first method utilizes the extrapolation of the real-axis integration accompanied with prelocating the singularities near the integration path and subtracting the contribution of the singularities from integrands. The second method performs deformed vertical integration path to integrate exponentially decay integrands. These two methods can be adaptively applied to different propagation distance and result in efficent computation of a wide range of fields. 
\par
As an application demonstration of our proposed formulation. The spatially dispersive graphene biased by drift current integrated with external photonic structures of uniaxially epsilon-near-zero hexagonal Boron Nitride is studied. Our computational results postulate that the electromagnetic response of graphene biased by drift current can be significantly reinforced when placed above epsilon-near-zero substrate and still preserves the unidirectional propagation properties.
\par
\section{Theory and Methodology}
\subsection{Problem defination and transmission line analogy formulation}
\begin{figure}[b]
	\includegraphics[width=\columnwidth]{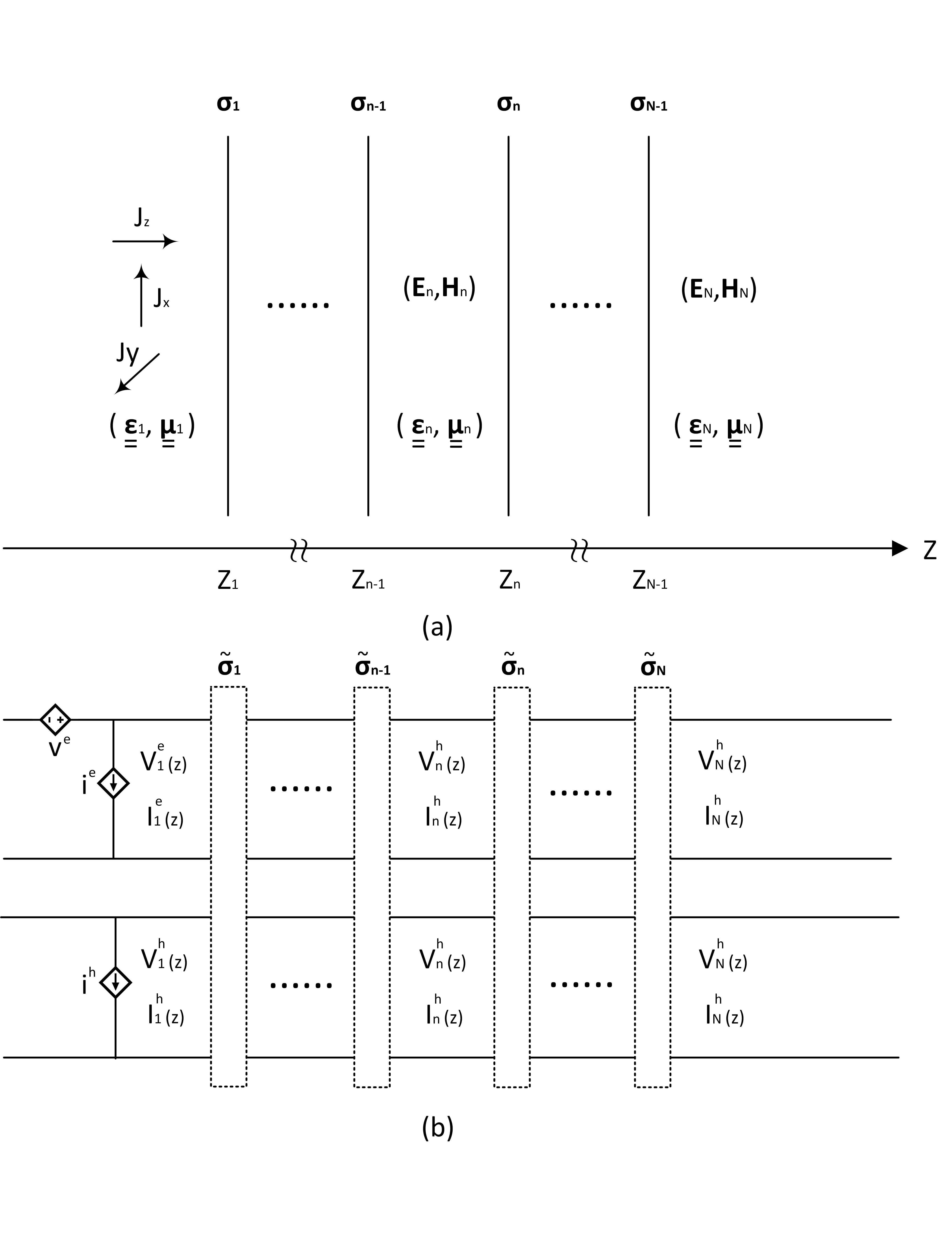}
	\caption{(a) The diagram of layered uniaxial media comprised of anisotropic conductive surfaces. (b) The spectral-domain transmisson line analog.}
	\label{fig:tl}
\end{figure}
We consider a unit-strength vertical or horizontal electric diploe located at $z'$ above the layered uniaxial media containing multiple in-plane anisotropic conductive sheets at the interfaces of the layers, which is shown in Fig. 1 (a). The structure is assumed to be of infinite lateral extent on transverse xy-plane and invariant in any plane transverse to the z-axis. The layered media may be uniaxially anisotropic, and the relative permittivity and permeability dyadic within the n-th layer are characterized by
\begin{equation}
{\underline{\underline{\bm{\varepsilon}_n}}}=\begin{bmatrix}
\epsilon_{tn} & 0 & 0\\
0 & \epsilon_{tn} & 0\\
0 & 0 & \epsilon_{zn}
\end{bmatrix}\,,\,{\underline{\underline{\bm{\mu}_n}}}=\begin{bmatrix}
\mu_{tn} & 0 & 0\\
0 & \mu_{tn} & 0\\
0 & 0 & \mu_{zn}
\end{bmatrix},
\label{eq1}
\end{equation}
where we also define 
\begin{equation}
\nu_n^e = \frac{\epsilon_{zn}}{\epsilon_{tn}}\,,\, \nu_n^h=\frac{\mu_{zn}}{\mu_{tn}}\,.
\end{equation}
The free-space wavenumber and intrinsic impedance will be denoted by $k_0$ and $\eta_0$ respectively, and the n-th layer transverse wavenumber is given as $k_{n}=k_0\sqrt{\varepsilon_{tn}\mu_{tn}}\,$. 
The conductive surfaces may be in-plane anisotropic, with a surface conductivity tensor
\begin{equation}
\dy{\bm{\sigma}}
=
\begin{bmatrix}
\sigma_{xx} & \sigma_{xy}\\
\sigma_{yx} & \sigma_{yy}
\end{bmatrix}
\label{sigxy}.
\end{equation}
To compute the dyadic Green function, the analysis is much facilitated by the Fourier transformation
of all transverse fields, which is defined as
\begin{equation}
\tilde{f}(\bm{k}_\rho)=
\int_{-\infty}^{\infty}\int_{-\infty}^{\infty}
f(\bm{\rho}) \,e^{j\bm{k}_\rho\cdot\bm{\rho}} \,dx dy\,,
\label{e2}
\end{equation}
where $\bm{\rho}=\hat{\bm{x}} x+\hat{\bm{y}} y$ and $\bm{k}_\rho=\hat{\bm{x}} k_x+\hat{\bm{y}} k_y$ is the transverse wavevector.
To facilitate the derivation, the field components in rotated coordinate are expressed as 
$\bm{\hat{u}}$, $\bm{\hat{v}}$ and $\bm{\hat{z}}$, where
\begin{equation}
\hat{\bm{u}} = \frac{\bm{k}_\rho}{k_\rho}\,,\,  \hat{\bm{v}} =\frac{\hat{\bm{z}}\times\bm{k}_\rho}{k_\rho}\,.
\end{equation}
The sigma tensor in rotated coordinate system is then transformed as
\begin{subequations}
\begin{eqnarray}
\dy{\tilde{\sigma}}
=
\begin{bmatrix}
\sigma_{uu} & \sigma_{uv}\\
\sigma_{vu} & \sigma_{vv}
\end{bmatrix}
=\dy{M}^T
\begin{bmatrix}
\sigma_{xx} & \sigma_{xy}\\
\sigma_{yx} & \sigma_{yy}
\end{bmatrix}
\dy{M}\,,
\end{eqnarray}
\begin{eqnarray}
\dy{M}=\frac{1}{k_\rho}
\begin{bmatrix}
k_x & -k_y\\
k_y & k_x
\end{bmatrix}\,.
\end{eqnarray}	
\end{subequations}
\par
 If Eq.~(\ref{e2}) is applied to the Maxwell's equations and note that the operator transforms as $\tilde{\bm{\nabla}} = -j\bm{k}_\rho + \hat{\bm{z}}\, d/d z$ \cite{michalski2005electromagnetic}. We can express the fields of the spectral-domain Green function inside the n-th layer as the following transmission line (TL) analog equations, which is illstrated by Fig. 1 (b) \cite{michalski2005electromagnetic}
\begin{subequations}
	\begin{gather}
	\frac{dV_n^p}{dz}(z,k_\rho)=-jk_{zn}^pZ_n^pI_n^p + v^p\,,\\
	\frac{dI_n^p}{dz}(z,k_\rho)=-jk_{zn}^pY_n^pV_n^p + i^p\,,\\
	k_{zn}^p = \pm\sqrt{{k_n}^2-{k_\rho}^2/\nu_n^p}\,,\,k_\rho^2 = k_x^2 + k_y^2\,,\\
	Z_n^e =\displaystyle\frac{1}{Y_n^e}=\eta_0\frac{k_{zn}^e}{k_0\epsilon_{tn}}\,, Z_n^h=\displaystyle\frac{1}{Y_n^h}=\eta_0\frac{k_0\mu_{tn}}{k_{zn}^h}\,,\\
	v^e = \displaystyle\frac{\eta_0k_\rho}{k_0\epsilon_{zn}} \hat{J}_z - \hat{M}_v\,,\, v^h = \hat{M}_u\,, \\
	i^e = -\hat{J}_u\,,\, i^h =-\displaystyle\frac{k_\rho}{k_0\eta_0\mu_{zn}}\hat{M}_z - \hat{J}_v\,.
	\end{gather}\label{eqtl}
\end{subequations}
where $V_n^p$ and $\ I_n^p$ denote the TL fields, $i^p$ and $\ v^p$ denote the sources. $k_{zn}^p$ and $ Z_n^p = \frac{1}{Y_n^p}$ are the charactristic paramters,  and the superscript $p = e,\,h\,$ indicates two independent modes of fields, the first being transverse-magnetic (TM) and the second transverse-electric (TE) to z-axis. The electric and magnetic fields of the n-th layer are then represented by the TL fields as
\begin{subequations}\label{eqf}
	\begin{eqnarray}
	\tilde{\bm{E}}_n \left( k_\rho,z\right)  = V_n^e\bm{\hat{u}} + V_n^h \bm{\hat{v}} - \displaystyle\frac{k_\rho \eta_0}{k_0 \epsilon_{zn} }I_n^e\bm{\hat{z}}\,,
	\end{eqnarray}
	\begin{eqnarray}
	\tilde{\bm{H}}_n \left(k_\rho,z\right)=-I_n^h\bm{\hat{u}}+I_n^e\bm{\hat{v}}+\displaystyle\frac{k_\rho}{k_0\mu_{zn}\eta_0}V_n^h\bm{\hat{z}}\,.
	\end{eqnarray}
\end{subequations}
Upon combing Eq.~(\ref{eqtl}) with Eq.~(\ref{eqf}), the spectal-domain dyadic Green Funtions in rotated coordinate are expressed as 
\begin{eqnarray}\label{eqsg}
\tilde{\dy{G}}^E &&=
\begin{bmatrix}
\tilde{E}_u^u & \tilde{E}_u^v & \tilde{E}_u^z\\
\tilde{E}_v^u & \tilde{E}_v^v & \tilde{E}_v^z\\
\tilde{E}_z^u & \tilde{E}_z^v & \tilde{E}_z^z
\end{bmatrix}\notag \\
&&=\begin{bmatrix}
-V_{ie}^e & -V_{ih}^e & -\displaystyle\frac{\eta_0 k_\rho}{k_0 \epsilon_{zn}^{\prime}} V_{ve}^e\\[8pt]
-V_{ie}^h & -V_{ih}^h & -\displaystyle\frac{\eta_0 k_\rho}{k_0 \epsilon_{zn}^{\prime}} V_{ve}^h\\[8pt]
\displaystyle\frac{\eta_0 k_\rho}{k_0 \epsilon_{zn}}I_{ie}^e & \displaystyle\frac{\eta_0 k_\rho}{k_0 \epsilon_{zn}}I_{ih}^e & -\displaystyle\frac{\eta_0^2k_\rho^2}{k_0^2\epsilon_{zn} \epsilon_{zn}^{\prime}}I_{ve}^e
\end{bmatrix}\,.
\end{eqnarray}
In Eq.~(\ref{eqsg}),  the supscripts and subscripts of V and I denote the voltage or current excited on the e or h mode TL by a unit voltage source on the e or h mode TL, respectively. To obtain the spatial-domain Green, we first return to $\hat{\bm{x}}$ $\hat{\bm{y}}$ coordinate via transform matrix and then conduct inverse Fourier transform
\begin{subequations}
\begin{eqnarray}
{\dy{\bm{G}}}^E
=\begin{bmatrix}
E_x^x & E_x^y & E_x^z\\E_y^x & E_y^y & E_y^z\\
E_z^x & E_z^y & E_z^z
\end{bmatrix}=\mathcal{F}^{-1}\{{\dy{\bm{A}}}\  \tilde{\dy{\bm{G}}}^E {\dy{\bm{A}}}^T\}\,,
\end{eqnarray}
\begin{eqnarray}
{\dy{\bm{A}}} = \frac{1}{k_\rho}\begin{bmatrix} k_x & -k_y & 0 \\
k_y & k_x & 0 \\
0 & 0 & 1
\end{bmatrix}\,.
\end{eqnarray}
\end{subequations}
More explicitly, the electric fields excited by a unit vertical dipole located in the 1st layer are
\begin{subequations}
	\begin{equation}
	G_z^z(z) = \mathcal{F}^{-1}\{- \displaystyle\dfrac{\eta_0^2}{k_0^2 \epsilon_{1}^{z} \epsilon_{1}^{z\prime}}k_\rho^2 I_{ve}^e(z,z')\}\,,\end{equation}
	\begin{equation}
	G_x^z(z) = \mathcal{F}^{-1}\{\displaystyle\dfrac{\eta_0}{k_0 \epsilon_1^{z\prime}} [k_x V_{ve}^e(z,z')-k_yV_{ve}^h(z,z')]\}\,,\end{equation}
\begin{equation}
	G_y^z(z) = \mathcal{F}^{-1}\{\displaystyle\dfrac{\eta_0}{k_0 \epsilon_1^{z\prime}} [k_y V_{ve}^e(z,z')+k_xV_{ve}^h(z,z')]\}\,.
	\end{equation}
\end{subequations}
\par
\subsection{Spectral-domain Green funciton}
To obtain the solution of spectral-domain TL voltage and current, T matrix  or RT method may be used to solve the TL fields. We only describe the result of T matrix here, as the IFCs can be extracted from this formulation. The detailed derivation can be found in Appendix~\ref{a1t}.
By expressing the voltages and currents amplitudes within the n-th section of the TL network as forward- and backward-propagating waves vectors $\bm{V}^\gtrless_n$, impedance matrix $\bm{Z}_n$, and $\bm{P}^\gtrless_n$ propagation matrix. The forward or backward-propagating voltage between n-th layer and (n+1)-th layer then are associated as
\begin{equation}
\begin{bmatrix}
\bm{V}^>_n\\
\bm{V}^<_n
\end{bmatrix}
=
\underbrace{
	\begin{bmatrix}
	\bm{\alpha}_n & \bm{\gamma}_n\\
	\bm{\delta}_n & \bm{\beta}_n
	\end{bmatrix}
}_{\bm{T}_{n}}
\begin{bmatrix}
\bm{V}^>_{n+1}\\
\bm{V}^<_{n+1}
\end{bmatrix}
\,,
\label{tjun}
\end{equation}
with
\begin{subequations}
	\begin{eqnarray}
	\bm{\alpha}_n=\frac{1}{2}\,
	\bm{P}_n^{-1}\big[\bm{1}+\bm{Z}_n\big(\bm{Y}_{n+1}+\tilde{\bm{\sigma}}_n\big)\big],
	\label{alpha}
	\end{eqnarray}
	\begin{eqnarray}
	\bm{\delta}_n=\frac{1}{2}
	\big[\bm{1}-\bm{Z}_n\big(\bm{Y}_{n+1}+\tilde{\bm{\sigma}}_n\big)\big],
	\end{eqnarray}
	\begin{eqnarray}
	\bm{\gamma}_n=\frac{1}{2}\,
	\bm{P}_n^{-1}\big[\bm{1}-\bm{Z}_n\big(\bm{Y}_{n+1}-\tilde{\bm{\sigma}}_n\big)\big]\bm{P}_{n+1},
	\end{eqnarray}
	\begin{eqnarray}
	\bm{\beta}_n=\frac{1}{2}
	\big[\bm{1}+\bm{Z}_n\big(\bm{Y}_{n+1}-\tilde{\bm{\sigma}}_n\big)\big]\bm{P}_{n+1}.\label{pn}
	\end{eqnarray}
	\label{eqabcd}
\end{subequations}
The above T matrices are $4 \times 4$ and contain $2 \times 2$ submatrices elements since they must incorporate the coupling between e and h modes. In any internal layer $n$, the amplitudes of the forward- and backward-propagating waves
are found as
\begin{eqnarray}
\begin{bmatrix}
\bm{V}^>_{n}\\
\bm{V}^<_{n}
\end{bmatrix}
=
\bm{T}_n\cdots\bm{T}_{N-2}
\bm{T}_{N-1}
\begin{bmatrix}
\bm{V}^>_N\\
\bm{V}^<_N
\end{bmatrix}
\,,
\label{anylayer}
\end{eqnarray}
where $n=N\!-\!1,N\!-2\!,\dots,1$. The outgoing wave amplitudes in the first and last layers are given as
\begin{equation}
\bm{V}^<_1=\overrightarrow{\bm{\Gamma}}\,\bm{V}^>_1\,,\quad
\bm{V}^>_N=\overrightarrow{\bm{\tau}}\, \bm{V}^>_1\,,
\label{ext}
\end{equation}
where the sources are represented as incident fields and enforced on the 1st layer
\begin{equation}
\bm{V}^>_1 = 
\left\{ \begin{array}{cccc}
\begin{bmatrix}
\frac{1}{2}\\0
\end{bmatrix}& if\ v^e, &
\begin{bmatrix}
\frac{Z_1^e}{2}\\0
\end{bmatrix}& if\ i^e, \\ \\
\begin{bmatrix}
0\\\frac{1}{2}
\end{bmatrix}& if\ v^h, &
\begin{bmatrix}
0\\\frac{Z_1^h}{2}
\end{bmatrix}& if\ i^h. 
\end{array} \right. 
\end{equation}
The total reflectance and transmittance coefficients are
\begin{eqnarray}
\overrightarrow{\bm{\tau}}
&&=
\begin{bmatrix}
\overrightarrow{\tau}^{ee} & \overrightarrow{\tau}^{eh} \notag \\
\overrightarrow{\tau}^{he} & \overrightarrow{\tau}^{hh}
\end{bmatrix}
=\bm{\alpha}^{-1}\\
&&= \frac{1}{det(\alpha)} \begin{bmatrix}
\alpha_{22} & -\alpha_{12}\\
-\alpha_{21} & \alpha_{11}
\end{bmatrix} = \frac{\bm{\mathcal{A}}}{\mathcal{D}} \,,
\label{eqtau}
\end{eqnarray}
\begin{eqnarray}
\overrightarrow{\bm{\Gamma}}
&& =
\begin{bmatrix}
\overrightarrow{\Gamma}^{ee} & \overrightarrow{\Gamma}^{eh}\\
\overrightarrow{\Gamma}^{he} & \overrightarrow{\Gamma}^{hh}
\end{bmatrix}
=\bm{\delta}\,\overrightarrow{\bm{\tau}} = \dfrac{\bm{\delta}\bm{\mathcal{A}}}{\mathcal{D}}\,.
\label{eqref}
\end{eqnarray}
The zeros of $\mathcal{D}$  represent the dispersion relation or IFCs of the system. To eliminate the singularities due to branch points, the equation below should be used instead and results in a pole-free  formulation \cite{michalski2018computation}
\begin{equation}
\mathcal{D}^\prime = Z^h_1 Z_N^e \mathcal{D}\,,
\end{equation}
on which the roots finding can be conducted to find all the zeros that represent the modes of SPWs.
\par
Although pole-free IFCs equation can be obtained from the T matrix method, it is numerically instable for the computation of the Green function. As $\bm{P}_{n}$ will be rounded to zero if $k^p_{zn} d_n$ is sufficiently large. This will arise issues when studying the surface wave propagating in structure comprised of thick layer. As a remedy, RT method can be utilized instead \cite{michalski2019modal,gu2021giant}. In our implementation, RT method is used for Green function computation and T matrix method is used for singularities tracking.
\par
\subsection{Spatial-domain Green funciton evaluation}
To obtain the spatial-domain Green Function, two-dimensional inverse Fourier transform needs to be computed. A direct  evaluation involving integration path along real axis can be computationally expensive especially for large radial distance. This is due to the factor that the inner integrand posseses oscillation proportional to the propagation distance. Moreover, singularities representing surface wave modes may exist near real axis and will result in an inefficient numerical quadrature's evaluation. Some methods have been proposed to mitigate this problems. Ref.~\cite{gangaraj2016directive} considers the hyperbolic response graphene and replaces the $k_x$ inner integral with residue terms and approximates the outer integral by stationary phase method. Ref.~\cite{michalski2019computation} applies deformed vertical integration path wrapped around branch point and derives closed-form residue terms of the surface wave modes for the anisotropic graphene in free space. However, previous works only consider simple geometries and limit the observation distance in the near-field regime. This work poses a more versatile numerical framework to solve problems consisiting of complex structures, e.g. hybridized with pronounced spatial dispersion, hyperbolic response of anisotropic graphene, and uniaxially anisotropicity of substrates. Our method is also adaptive to a wide range of observation and source location and designed to be very efficent.
\begin{figure}[!h]
	\includegraphics[width=\columnwidth]{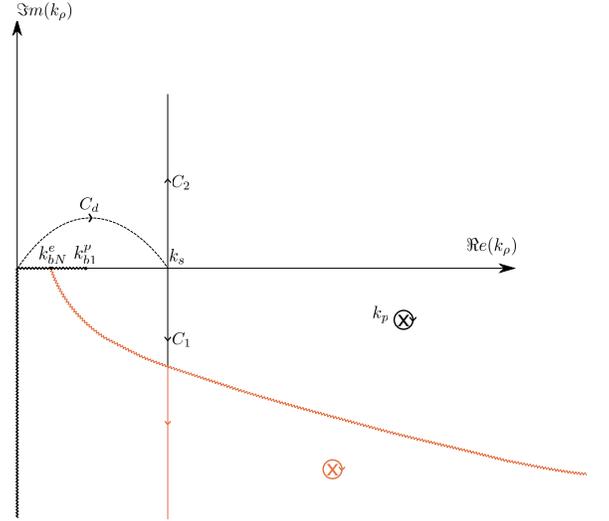}
	\caption{ The diagram of the complex-plane integration paths. $C_1$ and $C_2$ indicate vertical integration paths. The cross signs indicate the possible location of singularity points $k_p$. $k_{b1}^p$ is the Branch point of top layer, and its branch cut is indicated by the black zigzagging line. $k_{bN}^e$ is the TM mode branch point, and the branch cut of the epsilon-near-zero bottom layer is indicated by the orange zigzagging line. }
	\label{intpath}
\end{figure}
\par
We first represent the two-dimensional Fourier integral in cylindrical coordinate with $k_{\rho}$ and $ \xi$ as basis.
\begin{align}
G(\bm{\rho},z)=
\frac{1}{4\pi^2} \int_{0}^{2\pi}\int_{0}^{\infty}
\tilde{G}(\bm{k}_\rho,z)
e^{-j\bm{k}_\rho\cdot \bm{\rho} } k_\rho\,dk_\rho d\xi\,.
\label{e3}
\end{align}
When propagation distance $\hat{\bm{u}}\cdot \bm{\rho}$ is in the near and medium range regime, the real-axis integration path can be efficient if singularities near the integration path is properly extracted, so our first step is to develop a singularity substraction method to treat modal contribution of SPWs. 
\par As mentioned earlier, because the analytical form of singularities may not be attainable, numerical roots finding are required to track the SPWs modes. A robust method to conduct the roots finding in the complex plane involves using Cauchy argument principle to locate all the zeros inside specified contours \cite{michalski2018numerically}. This routine must be conducted for a series of discreated sampled $\xi$ value before conducting the numerical integration. Since $\mathcal{D^\prime}$ is multivalued in terms of $k_{z1}^p$ and $k_{zN}^p$, all Riemman sheets of $\mathcal{D^\prime}$ are piecewisely producted before conducting the root finding. Since the geometries considered in this paper only possess single SPWs mode, Muller's method described in Appendix~\ref{a2r} is used instead for faster computation, which typically requires as many as ten function evaluations to find the root.
Once the location of SPWs singularities $k_p$ is found, the residue can readily be found via computing 
\begin{equation}
r_p = \displaystyle\dfrac{\mathcal{N}(k_p)}{\frac{d}{d k_\rho}\mathcal{D}(k_\rho)|_{k_\rho = k_p}}\,,
\end{equation}
where $\frac{d}{d k_\rho}\mathcal{D}$ can be found by using the forward-mode auto differentiate and replacing the variables used in the spectral-domain Green function computation with dual number auto-differentiate variables \cite{michalski2018numerically}. The $\mathcal{N}$ and $\mathcal{D}$ are separated by Eq.~(\ref{eqref}). After the prelocation of singularities, the real-axis integration with the subtraction of SPWs singularities can be expressed as 
\begin{align}
G(\bm{\rho},z)=
\frac{1}{4\pi^2} \int_{0}^{2\pi} \bigg{\{} \int_{C_{d}} \tilde{G} e^{-j\bm{k}_\rho\cdot \bm{\rho} } k_\rho dk_\rho  + \int_{k_s}^{\infty} [\,\tilde{G}(\bm{k}_\rho,z) k_\rho \notag\\ - \sum\dfrac{r_p k_p}{k_\rho - k_p} e^{(k_\rho - k_p)(z+z^\prime)}\,]\,  
e^{-j\bm{k}_\rho\cdot \bm{\rho} }\,dk_\rho + \sum I_p\  \bigg{\}} \,d\xi\,,
\end{align}
where a detour path $C_d$ is taken between $k_\rho = 0$ and $k_s$, denoted by FIG. \ref{intpath} dashed line, to circumvent singularities due to branch points and the guided surface wave modes of layered media. The detour path is parameterized by 
\begin{subequations}
\begin{eqnarray}
	a = \frac{k_{s}}{2}\,,\,b = \frac{a}{4}\,,
\end{eqnarray}
\begin{eqnarray}
	k_\rho = s + jb \cos(\frac{\pi}{2a(s-a)})\,, s \in [0,k_s]\,,
\end{eqnarray}
\begin{eqnarray}
	dk_\rho = [1 -  \frac{jb\pi\sin(\frac{\pi}{2a(s-a)})}{2a} ]ds\,,
\end{eqnarray}
\begin{eqnarray}
k_s = 2 \max(\Re e(k_{bn}^p))\,,\,n = 1, N\,,
\end{eqnarray}
\end{subequations}
where $k_{bn}^p$ are the branch points of top and bottom layers. The plasmonic materials considered in this work may arise long-wavelength plasmonic modes. For passive materials, the SPWs singularities that need to be subtracted are located at fourth quadrant. The closed-form formula of $I_p$ can be found upon using the definition of the exponential integral function $E_1(\mathcal{Z})$ and noticing the possible enclosure of the singularity at origin \cite{gradshteyn2014table}.  
\begin{subequations}
\begin{eqnarray}
I_p = r_p k_p e^{-jk_pP}[E_1(\mathcal{P}+\mathcal{G}) - 2 \pi j H(P)]\,,
\end{eqnarray}
\begin{eqnarray}
P = \hat{\bm{u}}\cdot \bm{\rho}\,,
\end{eqnarray}
\begin{eqnarray}
\mathcal{P} = -jP(k_p-k_s)\,,\,\mathcal{G} = (k_p-k_s) (z+z^\prime)\,,
\end{eqnarray}
\end{subequations}
where $H(x)$ is the Heaviside step function. Notice we regularize the singularities terms by enforcing an exponentially decayed factor with the same ratio as the asymptotical behavior of the spectral-domain Green functions. The resulting tail integrand between $k_\rho = k_s$ to $\inf $ smoothly decays and can be evaluated efficently by applying low-order Gauss-Patterson quadrature between n consecutive break points separated by $q = \frac{\pi}{P}$, which gives a partial summation series $\{S_0^0,  ... , S_n^0\}$. The series convergence is then accelerated via Mosig-Michalski extrapolation algorithm \cite{patterson1968optimum,michalski2016efficient}. 
\begin{figure}[b]
	\includegraphics[width=\columnwidth]{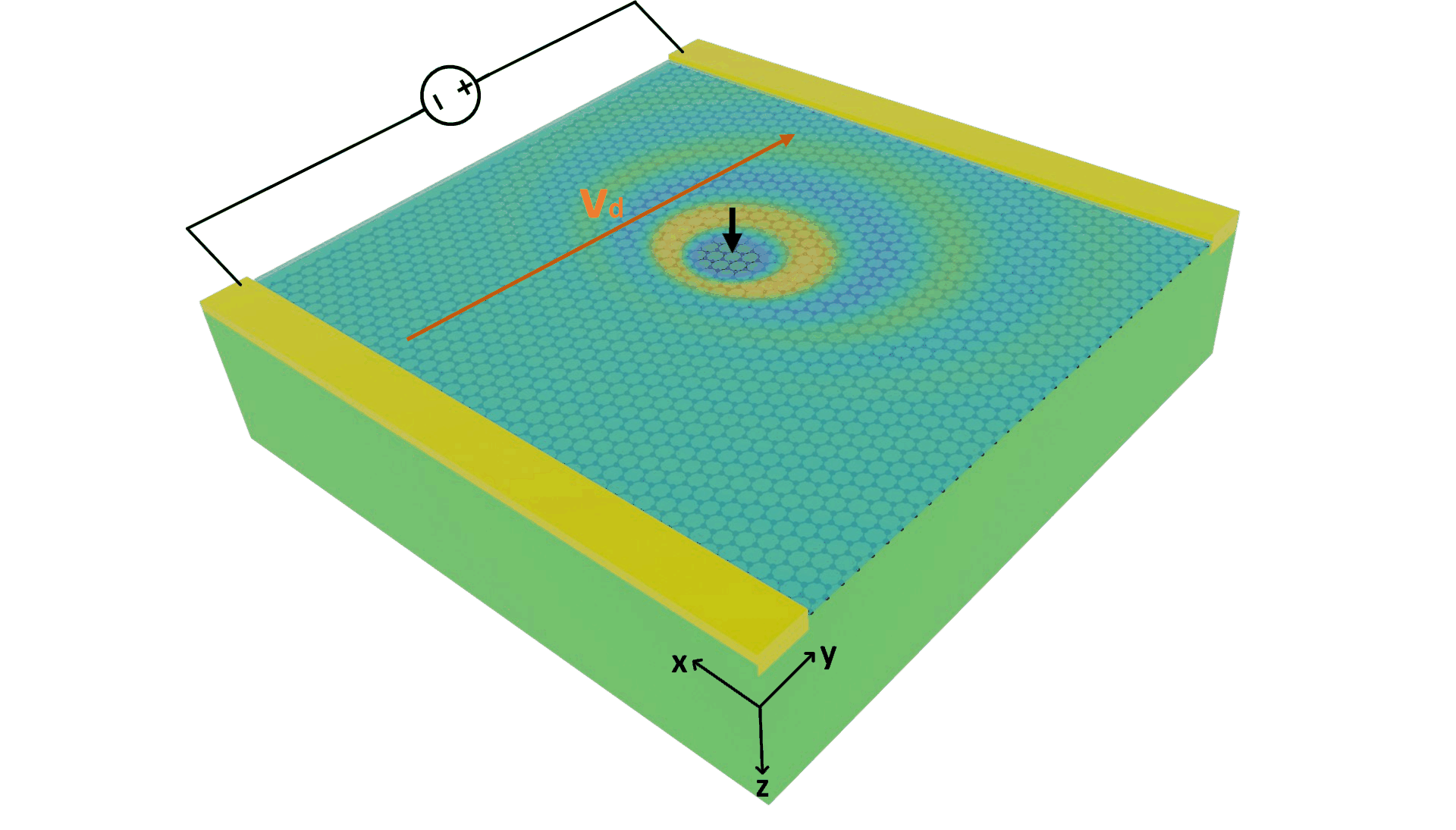}
	\caption{A two-layer structure with the graphene biased by drift current between the interface. The top layer is SiO$_2$, and the bottom layer is SiO$_2$ or uniaxial epsilon-near-zero hBN.  }
	\label{fig:geo}
\end{figure}
\begin{subequations}
\begin{equation}
	S_{n-k}^k = \displaystyle\dfrac{S_{n-k+1}^{k-1} - \eta_{n-k}^{k-1} S_{n-k}^{k-1}}{1 - \eta_{n-k}^{k-1}}\,,\,n \geq 1\,,\,1 \leq k \leq n\,,
\end{equation}
\begin{equation}
	\eta_n^k = \eta_n (\frac{x_n}{x_{n+1}})^{\mu k} \approx \dfrac{\eta_n}{1+\mu k q / x_n}\,,\,\mu = 2\,,
\end{equation}
\begin{equation}
\eta_n = \frac{\omega_{n}}{\omega_{n-1}} = - e^{-q\zeta}(\frac{x_n}{x_{n-1}})^\alpha\,,
\end{equation}
\end{subequations}
where the recursive scheme proceeds with $k=1,...,n$, and $S_0^n$ returns the best estimation of the series summation.  $\omega_n$ provides the structural information of the series by estimating the asymoptical behavior of the reminder. For example, for the inverse Fourier tranform of $\hat{G}_z^z$, we choose
\begin{eqnarray}
\alpha = 3,\ \zeta = z + z^\prime.
\end{eqnarray}
\par
 Another method to handle the singularities due to the SPWs modes is first traveling upwards in the first quadrant and then conduct the extrapolation of the integration along the real-axis direction, so that the SPWs singularities are kept away from the integrand \cite{michalski2015analysis}. This method is easiler to be implemented but takes slightly more computational costs. For far and moderate propagation distance, a deformed integration path can be performed, consisting of detour path, vertical path and residue, and we arrive at
\begin{align}
	G = \frac{1}{4\pi^2} \int_{0}^{2\pi}\bigg{\{} \int_{C_{d}} \tilde{G} e^{-j\bm{k}_\rho\cdot \bm{\rho} } k_\rho dk_\rho + H(P)2 \pi j  r_p  k_p e^{-j k_p P} \notag  \\
	  +\int_{0}^{\infty} \tilde{G} \,e^{-(jk_{s} + sgn(P) s^2) P} 2js[sgn(P)js^2-k_{s}] ds \bigg{\}} d\xi\,,
\end{align}
where the replacement $k_\rho = k_{s}- sgn(P) js^2$ is used in the first inner integral. If $\hat{\bm{u}}\cdot \bm{\rho} < 0$, $C_2$ path is taken. Otherwise $C_1$ path is taken and SPWs singularity will be enclosed. The branch cut line is defined as $\Im m (k_{zn}^{p}) = 0$.
For common isotropic materials, the branch cut is indicated by the black zigzagging line in FIG.\ref{intpath}. While in this work, the substrate with hyperbolic response and epsilon-near-zero properties $\Re e(\epsilon_t) \approx 0$ is also studied, which poses a special kind of branch cut line indicated by the orange zigzagging line in FIG.\ref{intpath} and can be described by
\begin{figure*}[!th]
	\centering
	\begin{minipage}{\columnwidth}
		\centering
		\includegraphics[width=0.72\columnwidth]{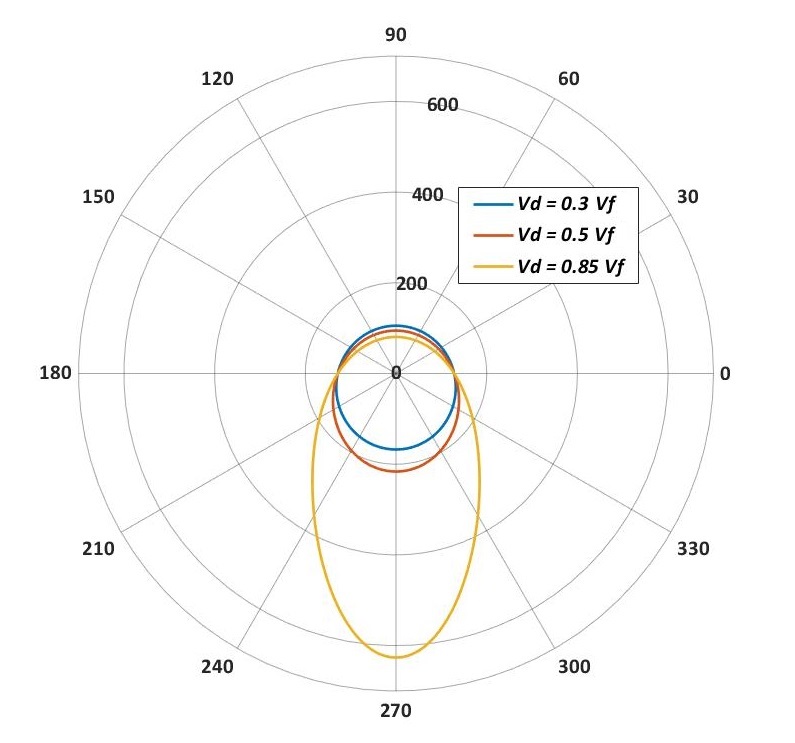}
		\caption*{(a)}
	\end{minipage}
	\begin{minipage}{\columnwidth}
		\centering
		\includegraphics[width=0.7\columnwidth]{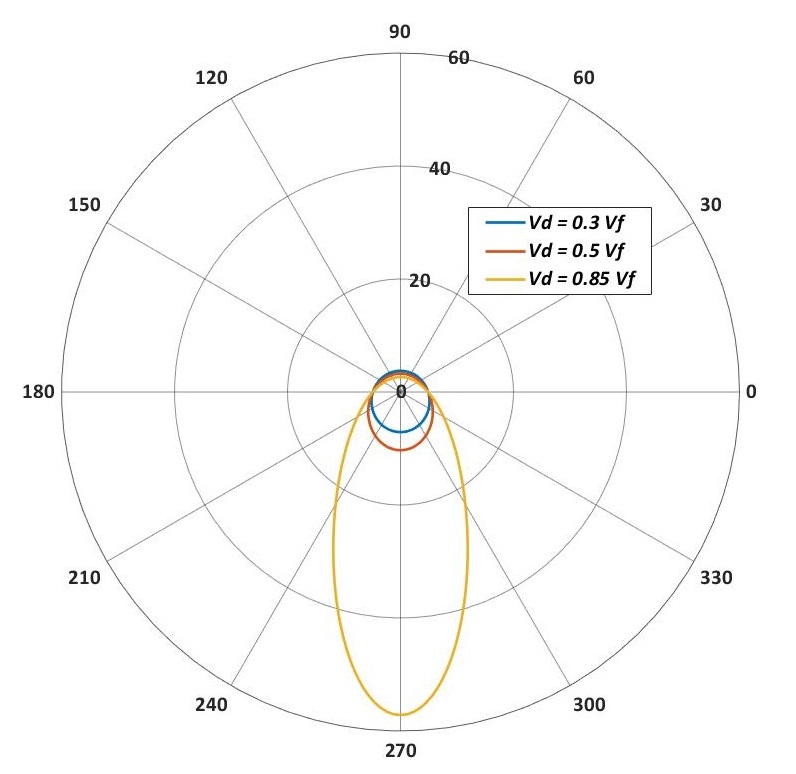}
		\caption*{(b)}
	\end{minipage}
\begin{minipage}{\columnwidth}
	\centering
	\includegraphics[width=0.72\columnwidth]{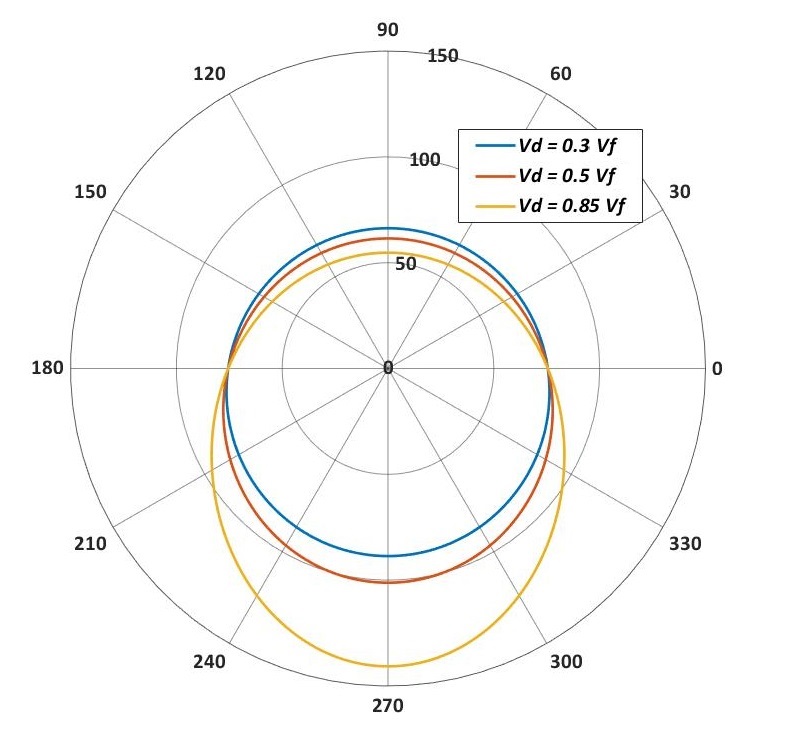}
	\caption*{(c)}
\end{minipage}
\begin{minipage}{\columnwidth}
	\centering
	\includegraphics[width=0.72\columnwidth]{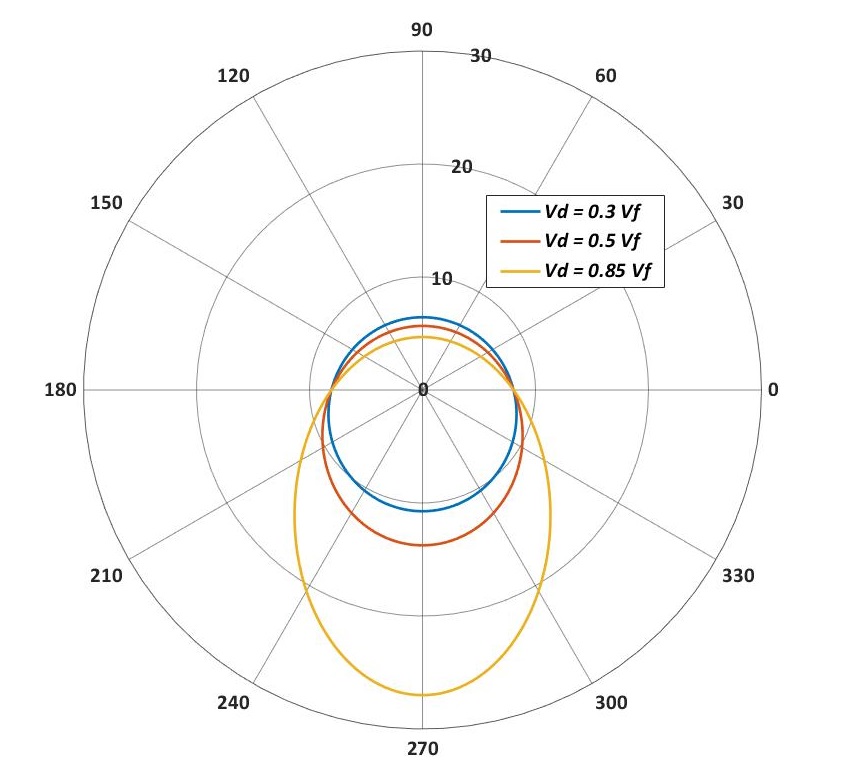}
	\caption*{(d)}
\end{minipage}
	\caption{ Real (a) and imagery (b) parts of IFCs for the graphene embeded in SiO$_2$ substrate. Real (c) and imagery (d) parts of IFCs for the graphene above epsilon-near-zero hBN substrate, with drift current velocities $ v_d = 0.3,\, 0.5,\, 0.85\, v_f$, respectively. }
	\label{figifc}
\end{figure*}
\begin{subequations}
 \begin{equation}
	k_i = \frac{k_r \Re e(\mathcal{V})+ \sqrt{|\mathcal{V}|^2 k_r^2-\Im m(\epsilon^t)\Im m (\mathcal{V})}}{\Im m(\mathcal{V})}\,,
 \end{equation}
  \begin{equation}
  k_r = \Re e(k_\rho)\,,\,k_i = \Im m (k_\rho)\,,\,\mathcal{V} = \frac{\epsilon^t}{\epsilon^z}\,.
   \end{equation}
\end{subequations}
The $C_1$ path will intecept hyperbolic materials' branch cut line and the sign of $k_{z}^e$ should be reversed once crossing the branch cut.
\par
 Finally, we point out that for extensively far field $\rho > 10 \lambda_0$, the $C_d$ detour integration is inefficent to be evaluated and needs to be replaced with multiply vertical integration paths wrapped around all branch points \cite{michalski2005electromagnetic}. Since this work focuses on the near and medium field range, this approach is not taken here. The outer integral has finite interval, globally adaptive numerical quadratures are applicable to its evaluation.
 \par
 \begin{figure*}
 	\centering
 	\begin{minipage}{\columnwidth}
 		\centering
 		\includegraphics[width=1.03\columnwidth]{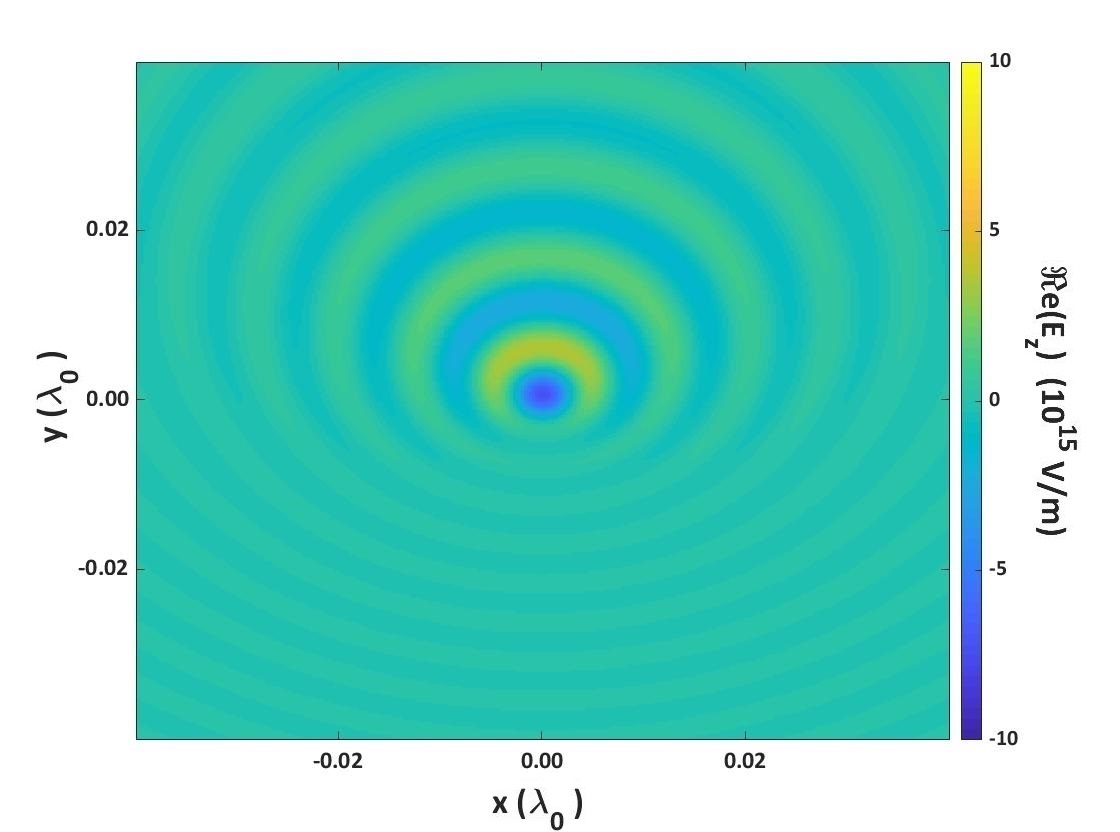}
 		\caption*{(a)}
 	\end{minipage}
 	\begin{minipage}{\columnwidth}
 		\centering
 		\includegraphics[width=1.02\columnwidth]{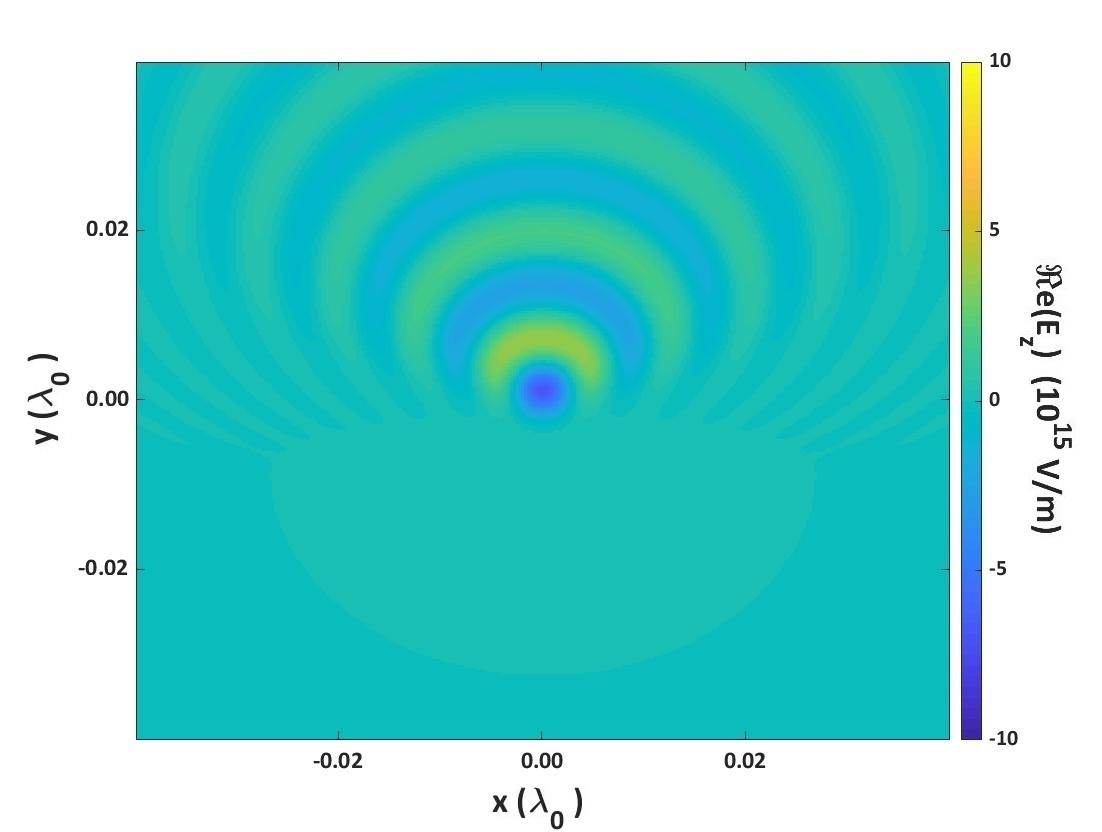}
 		\caption*{(b)}
 	\end{minipage}
 	\begin{minipage}{\columnwidth}
 		\centering
 		\includegraphics[width=1.02\columnwidth]{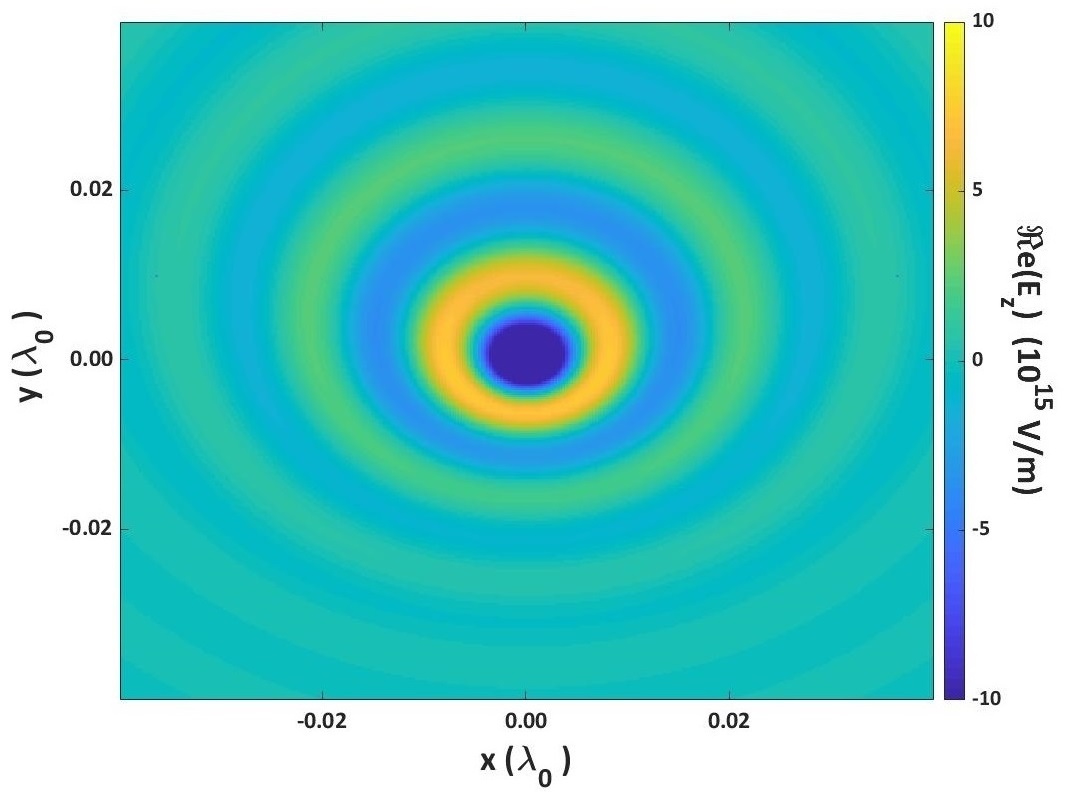}
 		\caption*{(c)}
 	\end{minipage}
 	\begin{minipage}{\columnwidth}
 		\centering
 		\includegraphics[width=1.025\columnwidth]{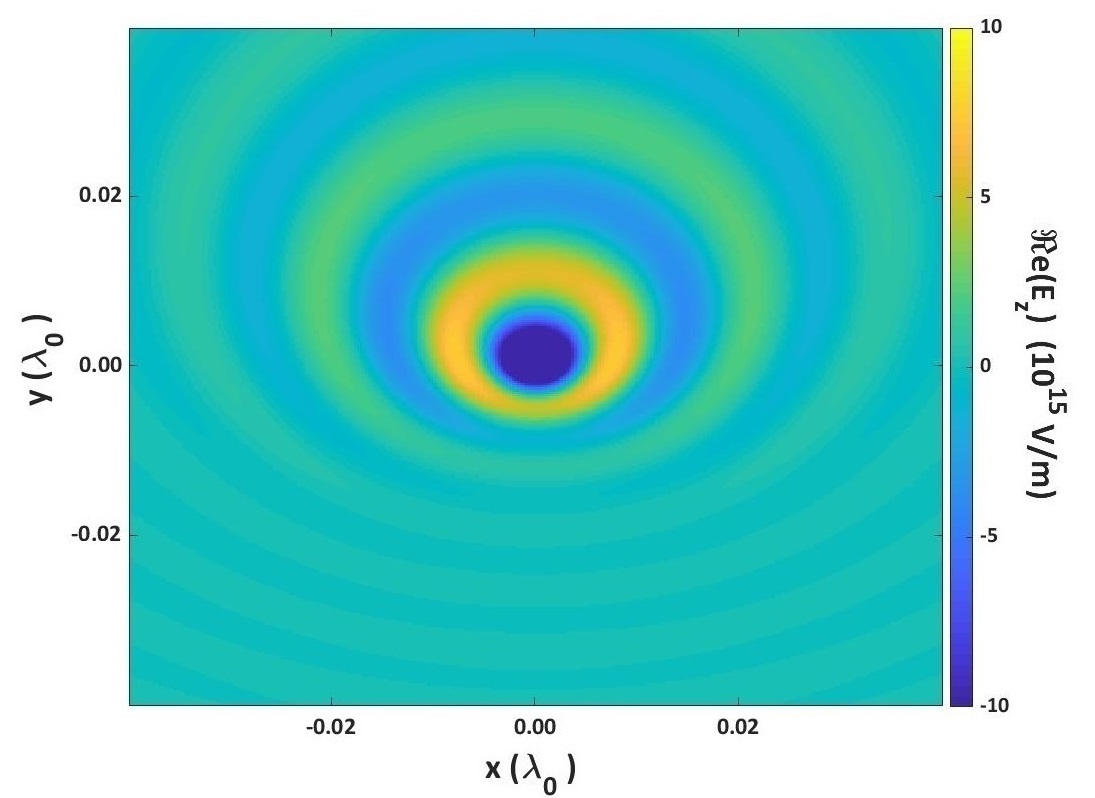}
 		\caption*{(d)}
 	\end{minipage}
 	\caption{xy-plane $\Re e(E_z)$ scattered fields of a vertical electric dipole located above the graphene embeded in SiO$_2$ substrate biased by drift current with velocities (a) $v_d = 0.5 \,v_f$, (b) $v_d = 0.85\, v_f$.  The graphene is placed above epsilon-near-zero hBN substrate with drift current velocities (c) $v_d = 0.5\, v_f$, (d) $v_d = 0.85\, v_f$.}
 	\label{figez}
 \end{figure*}
\section{Result}
\subsection{Elliptic spatially dispersive anisotropic graphene}
Recent advance in the field of graphene plasmonics shows that the graphene biased by drift current can break Lorentz reciprocity and support unidirectional surface waves \cite{correas2019nonreciprocal}. The surface conductivity of graphene exhibits spatial dispersion and anisotropicity. To model the macroscope response of graphene plasmons, the Bhatnagar-Gross-Krook formulation reported in \cite{lovat2013semiclassical} is applied to compute the graphene surface conductivity tensor $\dy{\bm{\sigma}}^{BGK}(\omega,k_x,k_y)\,$, with arbitrarily incident wave transverse wavevector. The conductivity formulation is multivaled, so before conducting the roots finding, two Riemman sheets are piecewisely producted. Also, for deformed path represention, two more vertical integration paths wrapped around the branch point are added Appendix~\ref{a3g}. When the drift current is applied to the graphene, the conductivity tensor will possess strong directivity and nonreciprocity, and can be modeled by the Doppler shift \cite{correas2019nonreciprocal}
\begin{equation}
\dy{\bm{\sigma}}
=
\frac{\omega}{\omega - k_y v_d}\dy{\bm{\sigma}}^{BGK}(\omega - k_y v_d,k_x,k_y)\,,
\end{equation}
where $v_d$ is the drift current velocities varies between $0.0\,-\,1.0\, v_f$. The first example we examinate is a graphene biased by  drift current embeded in Silicon dioxide (SiO$_2$, $\epsilon_r = 3.9$), which was considered in Ref.~\cite{morgado2020nonlocal}. Graphene parameters are $\mu_c = 0.05\, \textrm{eV},\, \tau = 500\, \text{ps}$, $\textrm{tempature}=300\,\textrm{K}$, and frequency of interest is $7.685\,\textrm{THz}$. The drift curent velocities are $ v_d = 0.3,\, 0.5,\, 0.85\, v_f$,  respectively, all pointing towards positive y-direction. The geometry is illustrated in FIG.\ref{fig:geo}.
\par
To understand the underlying modes of SPWs, the IFCs of the structure is first computed and shown in FIG. \ref{figifc} (a), (b), the IFCs are revealed to be a compressed ellipsoid pattern and $k_p$ possesses larger wavenumber close to negative y-direction. For the case of $ v_d = 0.5\, v_f $, SPWs wavelength is found to be $k_p / k_0 = 94.55 - 3.25j$ at $\xi = 90\deg$, which indicates along the drift current direction, the graphene surports short-distance SPWs. At $\xi = -90\deg$, the modal wavelength is $k_p / k_0 = 216.10 - 10.30j$.  The SPWs wavelength becomes much larger and exhibits unidirectional propagation \cite{wenger2018current}. It can be anticipated that SPWs propagting opposite to the drift current will be attenuated to zero within a very short distance. 
\par
\begin{figure}[ht]
	\centering
	\begin{minipage}{\columnwidth}
		\centering
		\includegraphics[width=1.05\columnwidth]{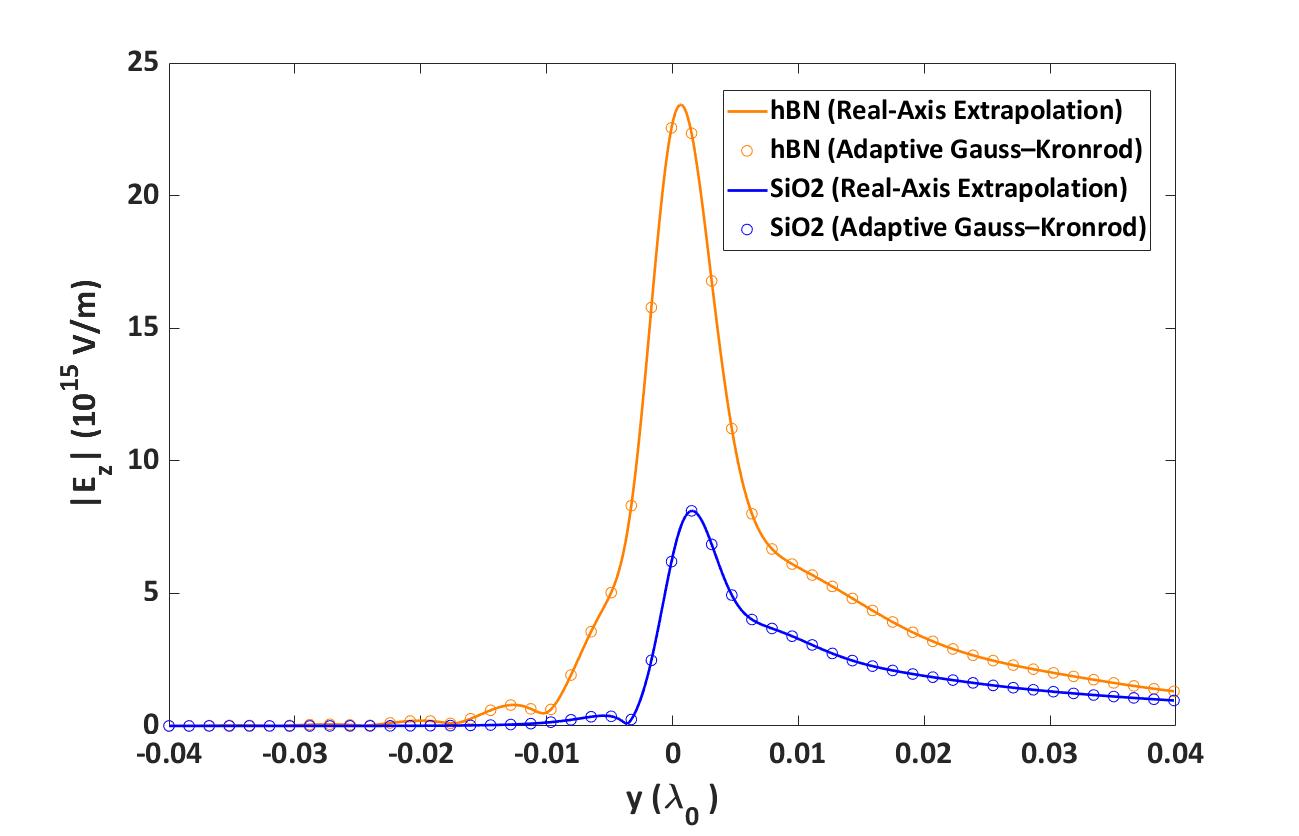}
		\caption*{(a)}
	\end{minipage}
	\begin{minipage}{\columnwidth}
		\centering
		\includegraphics[width=1.05\columnwidth]{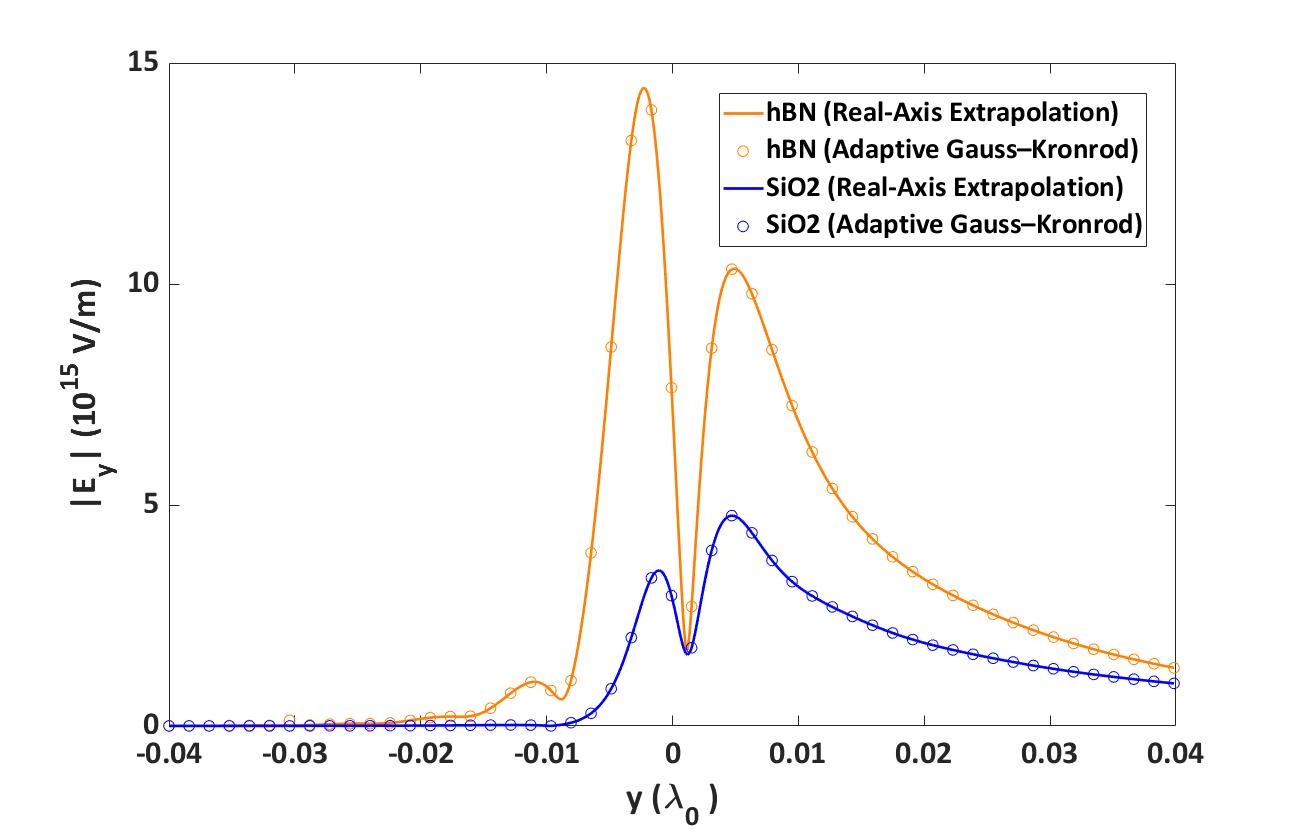}
		\caption*{(b)}
	\end{minipage}
	\caption{ Scattered fields (a) $|E_z|$, (b) $|E_y|$ along the y-axis, excited by a vertical electric dipole located above the graphene biased by $v_d = 0.85\, v_f $ drift current, with SiO$_2$ (blue line) or epsilon-near-zero hBN (orange line) as substrates. The dots data are computed using adaptive Gauss-Kronrod quadrature as a validation of our numerical method.}
	\label{figezycom}
\end{figure}
When the drift current is augmented to  $ v_d = 0.85 v_f $, the modal wavelength is found to be $k_p / k_0 = 80.61 - 2.65j$ at $\xi = 90\deg$, and $k_p / k_0 = 626.40 - 57.18j$ at $\xi = -90\deg$. A slight diminution of the SPWs wavelength and attenuation along the postive y-direction are observed due to an increased $ v_d$. While along the negative y-direction both the wavelength and attenuation increase dramstically. Those results postulate that the graphene biased by drift current surports unidirectional SPWs. As $v_d$ is increased to $1.0\,v_f$, SPWs propagating opposite to the drift current direction will be prevented \cite{morgado2020nonlocal}.
\par
Next, we examine a hybridized structure in which the graphene is placed above epsilon-near-zero (ENZ) hexagonal Boron Nitride (hBN) substrate. Previous studies demonstrate that ENZ materials possess unique properties such as supercoupling, emission enhancement, long-range SPWs mode \cite{silveirinha2006tunneling,enoch2002metamaterial,campione2015theory}. The effective wavelength of ENZ materials is much larger than the free space, and the waves propagating inside is spatially static. Those unconventional characteristics prompt us to hybridize ENZ materials with graphene biased by drift current to manipulate the plasmon-photon interaction.
\par
hBN is employed as a platform of the ENZ materials. The hBN is an uniaxially anisotropic and dispersive material, which supports two phonon-polariton modes in the THz regime that can exhibit zero or negative transverse epsilon \cite{dai2015graphene}. Lorentzian model described in Appendix~\ref{a4h} is applied to compute the epsilon tensor of hBN \cite{hajati2019modal}. The central frequency is chosen to be $7.685 THz$ which is near the type $\RomanNumeralCaps{1}$ phonon-polariton band of hBN, where the hBN transverse relative permittivity is $\epsilon_t \simeq 0.0$, and the longitudinal epsilon $\epsilon_z = 2.83$.
\par
FIG.\ref{figifc} (c) (d) show the IFCs of the graphene biased by drift current placed above the ENZ hBN substrate. For $ v_d = 0.5\, v_f $, the SPWs wavelength are $k_p / k_0 = 61.45 - 5.67j$ at $\xi = 90\deg$, and $k_p / k_0 = 101.20 - 13.75j$ at $\xi = -90\deg$. For  $ v_d = 0.85\, v_f $, we observe $k_p / k_0 = 54.72 - 4.686j$ at $\xi = 90\deg$, and $k_p / k_0 = 140.70 -17.01j$ at $\xi = 90\deg$. Compared with the case of employing SiO$_2$ as the substrate, the wavelength along positive y-axis is almostly halved, and the attenuation is slightly increased. Along the negative y-direction, the wavelength and attenuation are both decreased, but are still sufficient to prevent surface waves from prograting along negative y-direction. We thus conclude that utilizing ENZ materials as substrate will modify the SPWs wavelength towards a larger value.
\par
\begin{table*}
	\caption{\label{tab:t1}%
		The number of spectral-domain Green function evaluated in the Fourier integral
	}
	\begin{ruledtabular}\centering
		\begin{tabular}{cccc}
			\centering
			Radical distance ($\lambda_0$)&
			Real-axis integration with extrapolation\footnote{Note a., The number of Green function evaluated in $C_d$ integration + tail integration (number of partial summation terms used in extrapolation).}&
			Deformed vertical path\footnote{Note b., The number of Green function evaluated in $C_d$ integration + vertical path integration.}&
			Gauss–Kronrod\footnote{Note c., QUADPACK\cite{piessens2012quadpack} adaptive  Gauss–Kronrod quadrature routine DQAG is used.} \\
			\hline
			\colrule
			0.01 & 31+199(8) & 31+1215 & 841\\
			0.05 & 31+135(8) & 31+765 & 1591\\
			0.1 & 31+127(8) & 31+693 & 3181\\
			0.5 & 63+112(7) & 63+69 & 26553\\
			1.0 & 63+97(6) & 63+69 & 53643\\
			5.0 & 189+90(5) & 189+9 & 271989
		\end{tabular}
	\end{ruledtabular}
\end{table*}
Next, the scattered fields generated by a vertical unit electric dipole located $ 0.004\,\lambda_0$ above the graphene biased by drift current is analyzed, which provides a more illustrative picture. FIG.\ref{figez} shows the xy-plane $\Re e(E_z)$ fields where the observation height is fixed at $0.002\,\lambda_0$ above graphene. We observe, as expected, in all cases the electric fields are guided towards positive y-direction and decay rapidly along negative y-direction. For the case of $ v_d = 0.85\,v_f $ shown in Fig.\ref{figez} (b), the electric fields amplitude are almost zero for $y < 0$. Notably, a comparison of FIG.\ref{figez} (a) (b) with (c) (d) shows that a much higher field confinement above the graphene is achieved with ENZ hBN for both $v_d$ values. This is manifested more explicitly in FIG.\ref{figezycom} that plots the scattered fields $\Re e(E_z)$ and $\Re e(E_y)$ along the y-axis for a direct comparison of these two different substrates. The amplitude of electric fields in the vicinity of a dipole is more than three times larger for ENZ hBN and resembles the unidirectional fields pattern. Similar near-field enhancement due to ENZ material was also reported in dipole antennas array \cite{schulz2016optical}, the enhancement is attributed to the factor that, when a dipole source is placed in the vicinity of ENZ media, the radiation pattern will be predominantly backscattered normally towards the source media and thus results in strong backward emission with near-constant phase \cite{kim2016role}.
\par
\begin{figure*}
	\centering
	\begin{minipage}{\columnwidth}
		\centering
		\includegraphics[width=0.7\columnwidth]{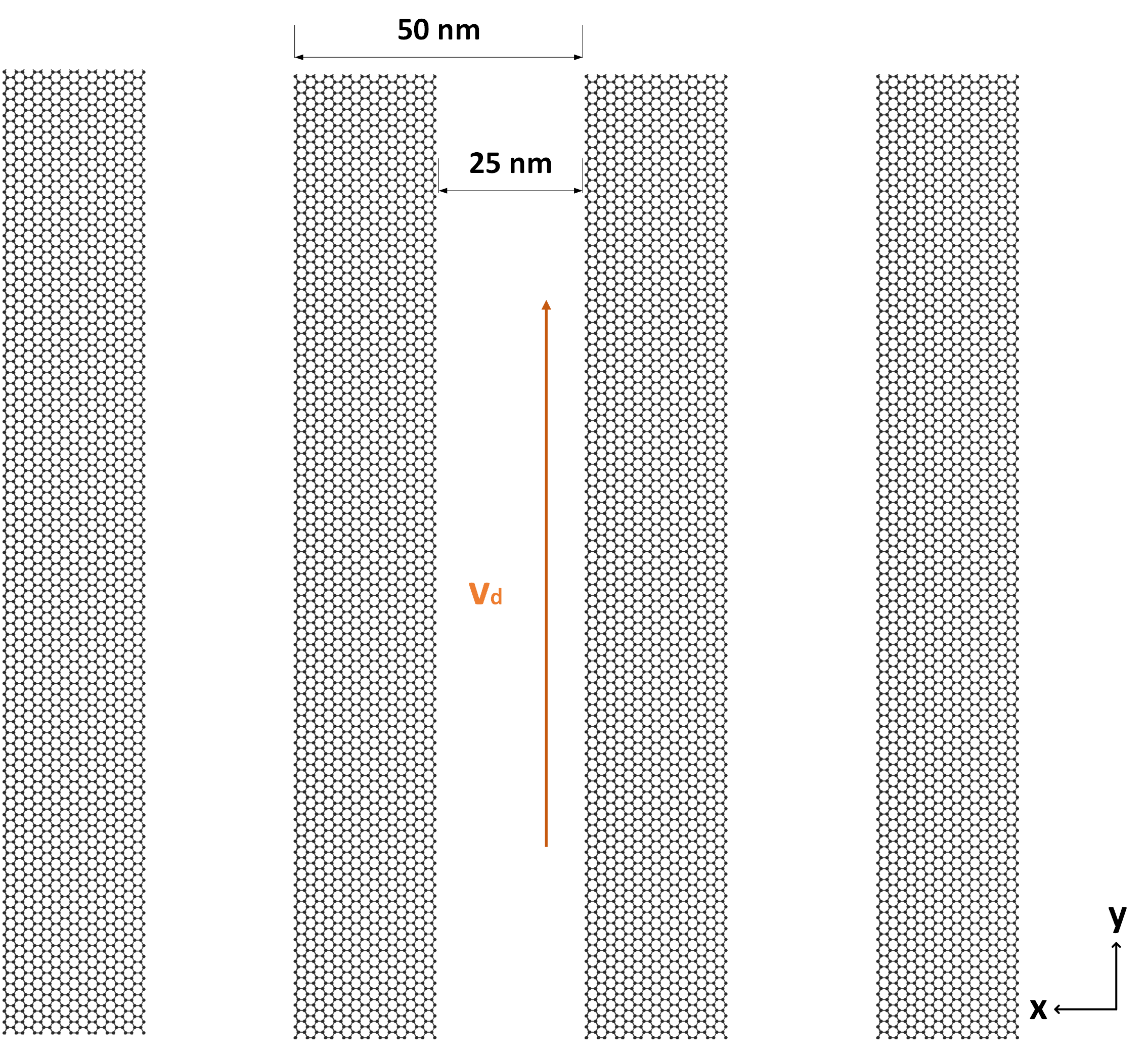}
		\caption*{(a)}
	\end{minipage}
	\begin{minipage}{\columnwidth}
		\centering
		\includegraphics[width=0.75\columnwidth]{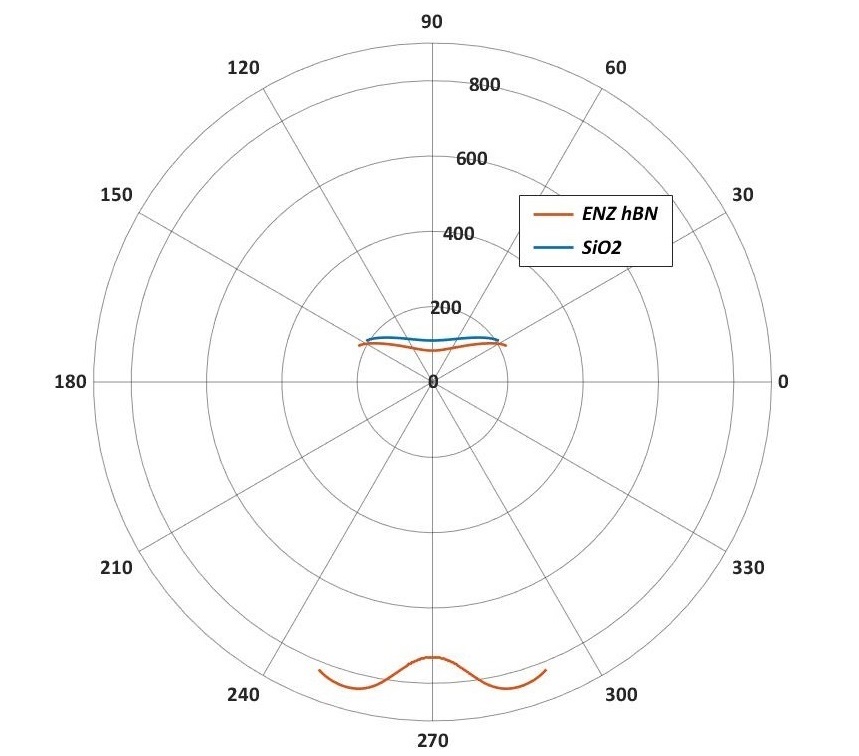}
		\caption*{(b)}
	\end{minipage}
	\begin{minipage}{\columnwidth}
		\centering
		\includegraphics[width=1.03\columnwidth]{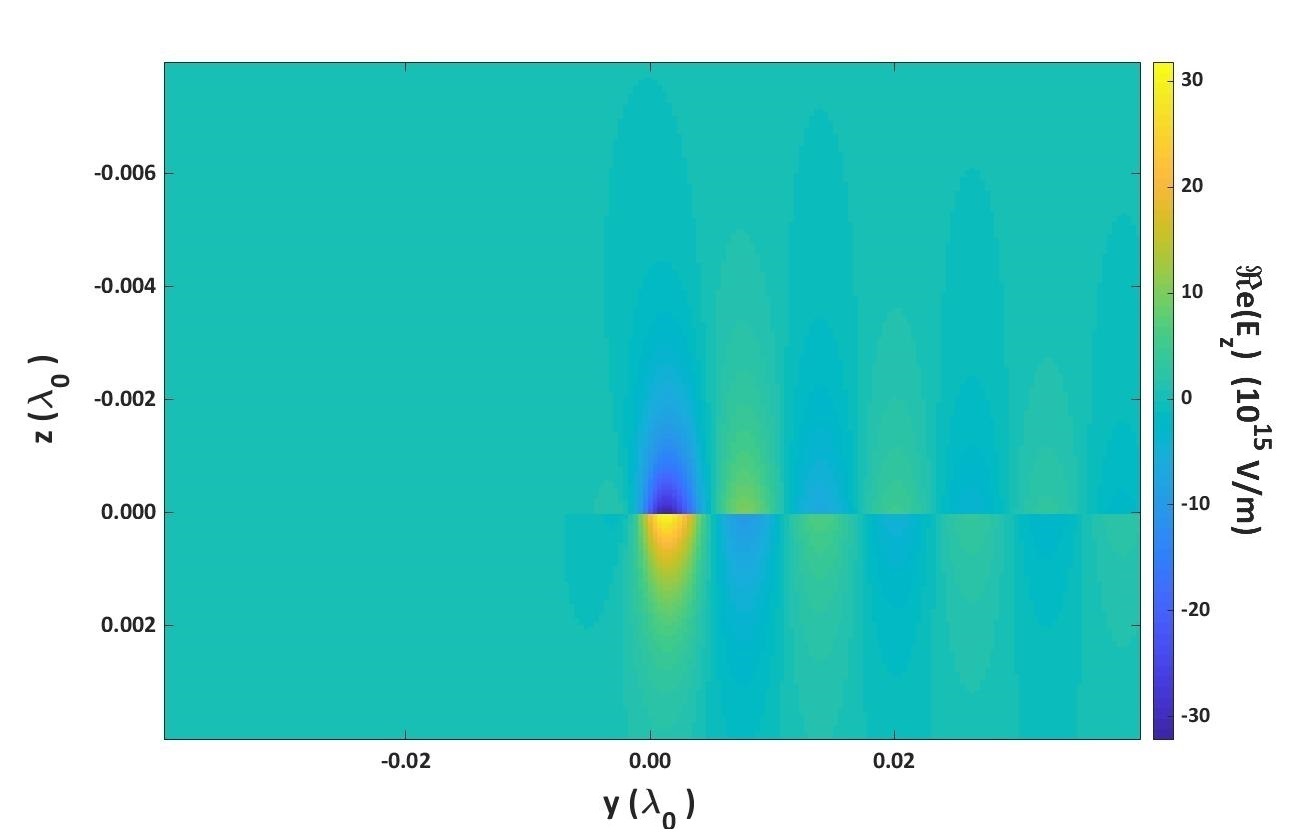}
		\caption*{(c)}
	\end{minipage}
	\begin{minipage}{\columnwidth}
		\centering
		\includegraphics[width=1.02\columnwidth]{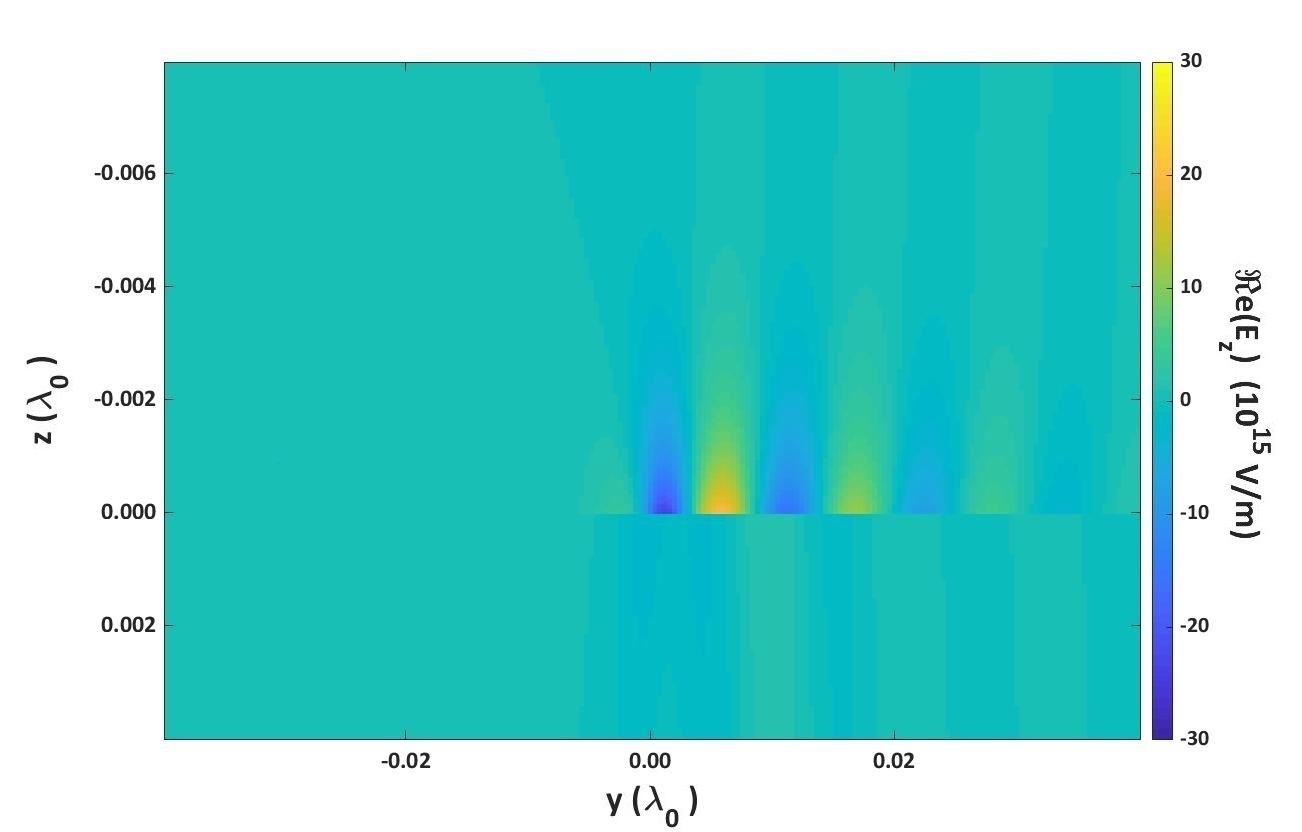}
		\caption*{(d)}
	\end{minipage}
	\begin{minipage}{\columnwidth}
		\centering
		\includegraphics[width=1.08\columnwidth]{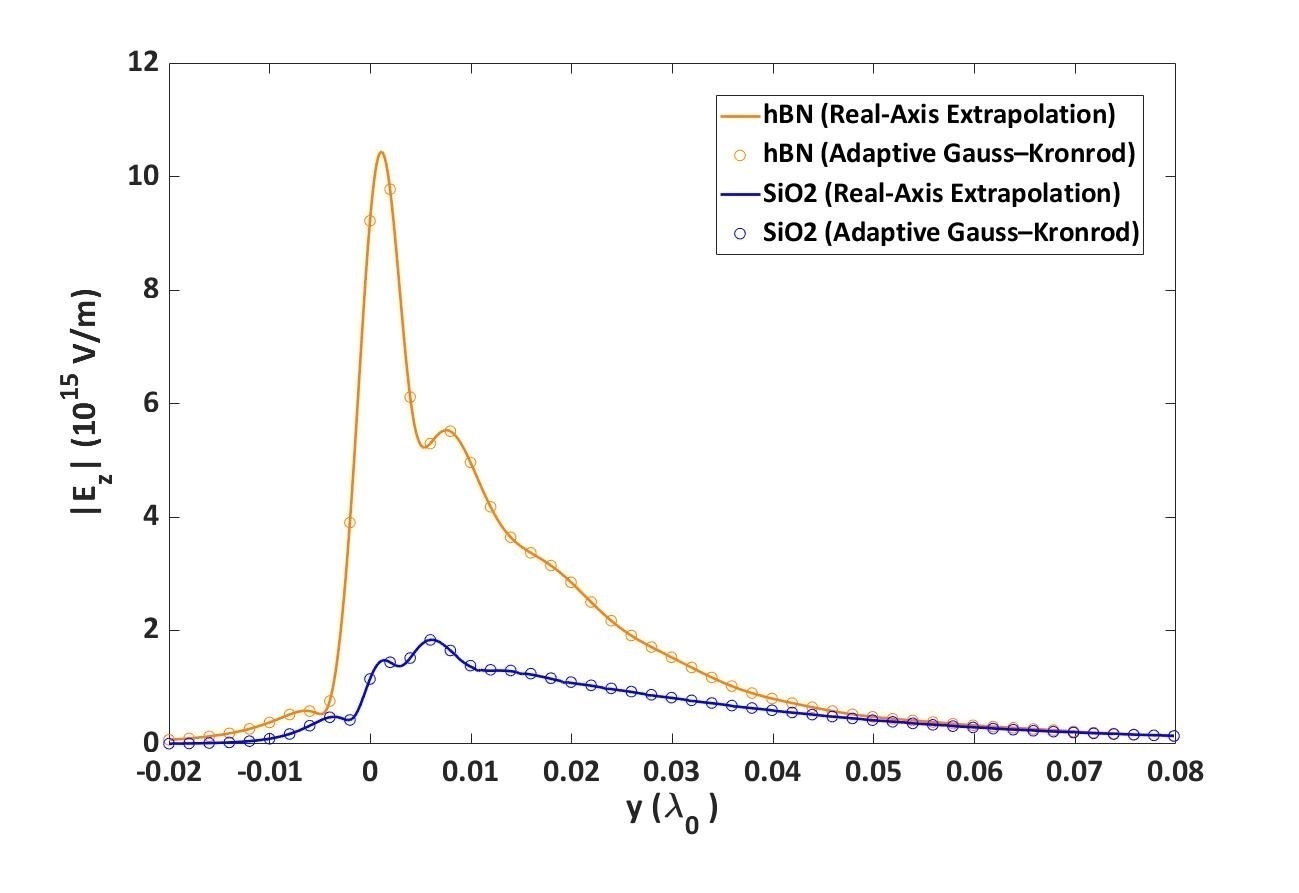}
		\caption*{(e)}
	\end{minipage}
	\begin{minipage}{\columnwidth}
		\centering
		\includegraphics[width=1.03\columnwidth]{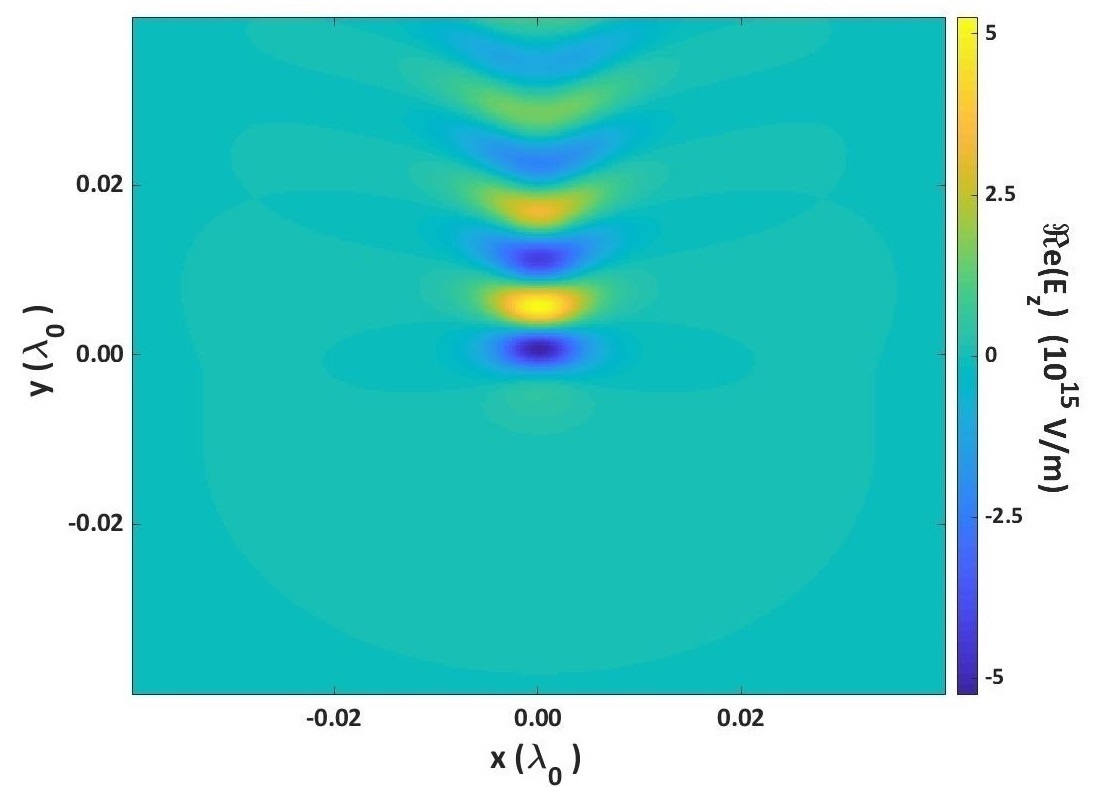}
		\caption*{(f)}
	\end{minipage}
	\caption{(a) An illustration of the densely periodic graphene ribbon array. (b) The real part of IFCs for the hyperbolic graphene above the SiO$_2$ and epsilon-near-zero hBN substrate, with $v_d = 0.85\, v_f$. (c) yz-plane scattered fields $\Re e(E_z)$ of SiO$_2$ as the substrate, and (d) ENZ hBN as the substrate. (e) The comparison of $|E_z|$ along the y-axis and data for validation. (f) xy-plane $\Re e(E_z)$ of the hyperbolic graphene above the epsilon-near-zero hBN substrate.}
	\label{figgra}
\end{figure*}
\subsection{Hyperbolic spatially dispersive anisotropic graphene}
\label{case2}
In the second case, a hyperbolic response conductive surface shown in FIG.\ref{figgra} (a) is considered. It was demonstrated that a densely-packed array of graphene ribbon array can possess effective surface conductivity tensor with $\textrm{sgn}(\Im m(\sigma_{xx})) \neq \textrm{sgn}(\Im m(\sigma_{yy}))$ and produce a hyperbolic response which arises extreme confinement of the supported SPWs and their channeling towards specific directions within the surface \cite{gomez2015hyperbolic,gomez2015hyperbolic2}. The ribbon array parameters are chosen as periodic length $L = 50\,\textrm{nm}$ and ribbon width $W=25\,\text{nm}$. The effective medium theory is applied to compute the surface conductivity \cite{luukkonen2008simple}.  Fig.\ref{figgra} (b) shows the IFC of the structure over different substrates. Notably, when employing SiO$_2$ as substate, no surface wave modes of $k_y < 0$ are supported. Along negative y-direction, ENZ hBN supports a series of modes possessing extremely large wavelength for $k_y < 0$ and $\Im m(k_p)\simeq 200 \,k_0$, which prevents SPWs from propagating along this direction. Along y-direction, the SPWs wavelength of ENZ hBN is only slightly smaller than the SiO$_2$ case, and the surported modes are limited within a cramped range, which implies the extreme fields confinement due to hyperbolic response. 
\par
The  $\Re e(E_z)$  of graphene ribbon array over the ENZ hBN substrate is computed to demonstrate the extremely strong light-matter interaction of the hybridized structure. We observe from FIG.\ref{figgra} (f) that SPWs induced by graphene ribbon array exhibit extreme fields confinement due to the hyperbolic plasmonic mode, and almost no surface waves propagating along negative y-direction. The comparison of $|E_z|$ amplitude in FIG.\ref{figgra} (e) indicates that over five times stronger fields are obtained when utilizing the ENZ hBN as substrate. 
\par
Fig.\ref{figgra} (c) (d) show the yz-plane $\Re e(E_y)$ of both substrates. Notice in Fig.\ref{figgra} (d), there is almost no electric fields inside the ENZ hBN underneath $z=0$, which leads to much stronger fields confinement above the graphene. ENZ hBN effectively reinforces the light-matter interaction of the graphene ribbon array, and preserve the nonreciprocity and hyperbolic response of the system.
\par
\subsection{Computational costs analysis}
 Finally a comparison of the results obtained by the proposed methods with the corresponding results generated by an adaptive Gaussian quadrature is presented. We conduct the benchmark based on the structure of section \ref{case2}. Since the evlaution of the inner integrand is the key of computational costs, the number of spectral-domain inner function evaluated during the inner Fourier integral at $\xi = 90\deg$ for various of radical distance is recorded in TABLE \ref{tab:t1}. The relative error is set to be $10^{-6}$ as convergence criteria.
\par
We observe the real-axis integration method accompanied with the extraplotion and singularity subtraction requires a less quantity of function evalutions for all fields range tested. The extrapolation rapidly converges within less than ten terms of partial summation. The deformed path requires less function evalution only within moderate and far range field distance, and is less efficient for near field due to the slow decay of the integrand. These two methods are both considerately more efficent than directly applying the globally adaptive quadrature which doesn't handle the ossciliation of the integrand. The singularity subtraction is found to be critcal for the convergence of extrapolation method, without which convergence criteria are never met and doesn't achieve any speed-up.
\par
\section{Conclusion}
Formulation and computational schemes of the surface plasmonic waves excited by dipole sources above anisotropic and spatially dispersive two-dimensional surfaces of infinite extent embedded in planarly layered uniaxial media are proposed. The spatial-domain Green function is rigorously solved by a two-dimensional Fourier integral. We also demonstrate that surface plasmon-photon waves of graphene exhibiting significant spatial dispersion can be substantially enhanced when hybridized with epsilon-near-zero substrate. As a possible future work, the dyadic Green function may be utilized as kernels in the integral equation method to model more complex geometries \cite{michalski1990electromagnetic}.

\appendix
\section{T Matrix Method} 
\label{a1t}
The T matrix method \cite{michalski2019modal,gu2021giant} is used to solve the iso-frequency contours and spectral-domain Green function.
To obtain the solution of spectral-domain TL voltage and current,
we first express the voltages and currents within the n-th section of the TL network as forward- and backward-propagating voltage waves represented by column vectors  $\bm{w}^\gtrless_n$, where $\bm{V}^\gtrless_n$ are amplitudes vectors, and $\bm{P}^\gtrless_n$ are propagation matrix.
\begin{subequations}
\begin{eqnarray}
\bm{w}^\gtrless_n(z)=
\begin{bmatrix}
w^{e\gtrless}_n(z)\\w^{h\gtrless}_n(z)
\end{bmatrix}\,,\,
\bm{V}^\gtrless_n=
\begin{bmatrix}
V^{e\gtrless}_n
V^{h\gtrless}_n
\end{bmatrix}\,,\,
\end{eqnarray}
\begin{eqnarray}
\bm{P}^\gtrless_n(z) =
\begin{bmatrix}
e^{\mp jk_{zn}^e(z-z_n^\lessgtr)}&0\\
0&e^{\mp jk_{zn}^h(z-z_n^\lessgtr)}
\end{bmatrix}\,,
\end{eqnarray}
\begin{eqnarray}
z_n^< = z_{n-1}\,,\,z_n^> = z_{n}\,,\,
\end{eqnarray}
\begin{eqnarray}
\begin{bmatrix}
\bm{w}^>_n(z)\\
\bm{w}^<_n(z)
\end{bmatrix}
=
\begin{bmatrix}
\bm{P}^>_n(z) & \bm{0}\\
\bm{0} & \bm{P}^<_n(z)
\end{bmatrix}
\begin{bmatrix}
\bm{V}^>_n\\
\bm{V}^<_n
\end{bmatrix}\,,\,
\end{eqnarray}
\begin{eqnarray}
\bm{P}_n =
\begin{bmatrix}
e^{-jk_{zn}^ed_n}&0\\
0&e^{-jk_{zn}^hd_n}
\end{bmatrix}.
\end{eqnarray}
\end{subequations}
The voltage and current in n-th layer can be expressed as
\begin{gather}
\begin{bmatrix}
\bm{V}_n(z)\\
\bm{I}_n(z)
\end{bmatrix}
=
\begin{bmatrix}
\bm{1} & \bm{1}\\
\bm{Y}_n & -\bm{Y}_n
\end{bmatrix}
\begin{bmatrix}
\bm{w}^>_n(z)\\
\bm{w}^<_n(z)
\end{bmatrix}\,,\,
\bm{Y}_n =
\begin{bmatrix}
Y^e_n&\bm{0}\\
\bm{0}&Y^h_n
\end{bmatrix},
\label{vinz}
\end{gather}
where $\bm{1}$ and $\bm{0}$ are $2\times2$ identity and zero matrices, respectively.
Notice the $1\times4$ vectors and $4\times4$ matrices of the above equations contain both e and h modes parameters. To enforce the boundary condition at the interfaces between adjacent layers,
where conductive sheets are present, a transformer between n-th and (n+1)-th layers is characterized by
\begin{equation}
\begin{bmatrix}
\bm{V}_n(z_n^>)\\
\bm{I}_n(z_n^>)
\end{bmatrix}
=
\begin{bmatrix}
\bm{1} & \bm{0}\\
\hat{\bm{\sigma}}_n & \bm{1}
\end{bmatrix}
\begin{bmatrix}
\bm{V}_{n+1}(z^>_n)\\
\bm{I}_{n+1}(z^>_n)
\end{bmatrix},
\label{bcvi}
\end{equation}
where  $\hat{\bm{\sigma}}_n$ is the transformed surface conductivity matrix. We thus
combine Eq.~(\ref{bcvi}) and Eq.~(\ref{vinz}) and arrive at
\begin{eqnarray}
\begin{bmatrix}
\bm{1} & \bm{1}\\
\bm{Y}_n & -\bm{Y}_n
\end{bmatrix}
\begin{bmatrix}
\bm{w}^>_n(z^>_n)\\
\bm{w}^<_n(z^>_n)
\end{bmatrix}
=\notag \\
\begin{bmatrix}
\bm{1} & \bm{1}\\
\hat{\bm{\sigma}}_n+\bm{Y}_{n+1} & \hat{\bm{\sigma}}_n-\bm{Y}_{n+1}
\end{bmatrix}
\begin{bmatrix}
\bm{w}^>_{n+1}(z^>_n)\\
\bm{w}^<_{n+1}(z^>_n)
\end{bmatrix}.
\label{tjump}
\end{eqnarray}
The forward or backward-propagating voltage between n-th layer and (n+1)-th layer then are associated as
\begin{equation}
\begin{bmatrix}
\bm{V}^>_n\\
\bm{V}^<_n
\end{bmatrix}
=
	\begin{bmatrix}
	\bm{\alpha}_n & \bm{\gamma}_n\\
	\bm{\delta}_n & \bm{\beta}_n
	\end{bmatrix}
\begin{bmatrix}
\bm{V}^>_{n+1}\\
\bm{V}^<_{n+1}
\end{bmatrix}
\,,
\label{tjun}
\end{equation}
the matrix elements of which are given in Eq.~(\ref{eqabcd}).
\section{Roots Finding for Iso-frequency Contours}
\label{a2r}
Muller's method \cite{press1986numerical} is used to conduct the zeros searching since the structure we consider in this work only has one surface plasmonic mode. Three initial points $z_0,z_1,z_2$ are chosen based on the estimated $k_p$ location, and the iterative starts from $i = 0\,$,
\begin{subequations}
	\begin{equation}
	h_0 = z_{i+1} - z_i\,,\, h_1 = z_{i+2} - z_{i+1}\,,\, 
	\end{equation}
	\begin{equation}
	d_0 = \frac{f(z_{i+1}) - f(z_{i})}{h_0}\,,\, d_1 = \frac{f(z_{i+2}) - f(z_{i+1})}{h_1}\,,\,
	\end{equation}
	\begin{equation}
	a = \frac{d_1-d_0}{h_1+h_0}\,,\,b=ah_1+d_1\,,\,c=f(z_{i+2})\,,
\end{equation}
\end{subequations}
\begin{subequations}
\begin{eqnarray}
	\,\mathcal{R} = \sqrt{b^2-4ac}\,,\,\\
	D = \begin{cases}
	b+\mathcal{R},& |b+\mathcal{R}| > |b-\mathcal{R}|\\
	b-\mathcal{R},& \text{otherwise}
	\end{cases}\,,
	\end{eqnarray}
\end{subequations}
\begin{eqnarray}
	\Delta z = \frac{-2c}{D}\,,\, z_r = z_{i+2} + \Delta z\,.
	\end{eqnarray}
The procedure stops if $\Delta z$ is small enough. Otherwise, before the next iterative starts, $z_{i+2} = z_r$. The method above is appropriate to the structures that excite single surface plasmonic modes. For multiply surface plasmonic modes, Cauchy argument principle should be used instead on the IFCs to find all the zeros inside 
predetermined contours on $k_\rho$ plane \cite{michalski2018numerically}.
\section{Graphene Surface Conductivity Tensor}
\label{a3g}
The surface conductivity tensor used in this paper is derived from the semiclassical Boltzmann transport equation under both the relaxation-time approximation and the Bhatnagar-Gross-Krook model, which models the intraband transitions of graphene and includes the spatial dispersion for transverse wavevector. The formulation becomes inapplicable when $|k_\rho|> 2\,k_f$. The closed-form  expressions below are strictly correct only when $\mu_c = 0\, \text{eV}$. However, numerical results confirm that the closed-form expressions results show a very good agreement with the exact numerical integration over the first Brillouin zone using the tight-binding electron dispersion relation \cite{lovat2013semiclassical}.
\begin{subequations}
	\begin{equation}
	\sigma_{xx}^{BGK} (k_x,k_y) = \gamma \dfrac{I_{\phi_{xx}} + \gamma_D \Delta k_y (I_{\phi_{xx}}k_y - I_{\phi_{yx}}k_x)}{D_{\sigma}}\,, 
	\end{equation}
	\begin{equation}
	\sigma_{xy}^{BGK} (k_x,k_y) = \gamma \dfrac{I_{\phi_{xy}} + \gamma_D \Delta k_y (I_{\phi_{xy}}k_y - I_{\phi_{yy}}k_x)}{D_{\sigma}}\,,
	\end{equation}
	\begin{equation}
	\sigma_{yx}^{BGK} (k_x,k_y) = \gamma \dfrac{I_{\phi_{yx}} + \gamma_D \Delta k_x (I_{\phi_{yx}}k_x - I_{\phi_{xx}}k_y)}{D_{\sigma}}\,,
	\end{equation}
	\begin{equation}
	\sigma_{yy}^{BGK} (k_x,k_y) = \gamma \dfrac{I_{\phi_{yy}} + \gamma_D \Delta k_x (I_{\phi_{yy}}k_x - I_{\phi_{xy}}k_y)}{D_{\sigma}}\,,
	\end{equation}
\end{subequations}
with
\begin{subequations}
	\begin{equation}
	I_{\phi_{xx}}(\omega,k_x,k_y) = 2\pi\dfrac{v_f^2 k_y^2 k_t^2 R - \alpha v_f k_x k_q^2 -\alpha^2 k_q^2(1-R)}{v_f^2(\alpha+v_fk_x)k_t^4}\,, \end{equation}
	\begin{eqnarray}
	I_{\phi_{xy}}(\omega,k_x,k_y) = I_{\phi_{yx}}(k_x,k_y) = \notag \\ -2\pi k_x k_y\dfrac{v_f^2 k_t^2 R + 2\alpha v_f k_x +2\alpha^2 (1-R)}{v_f^2(\alpha+v_fk_x)k_t^4}\,,
	\end{eqnarray}
	\begin{equation}
	I_{\phi_{yy}}(\omega,k_x,k_y) = 2\pi\dfrac{v_f^2 k_x^2 k_t^2 R + \alpha v_f k_x k_q^2 +\alpha^2 k_q^2(1-R)}{v_f^2(\alpha+v_fk_x)k_t^4}\,,
	\end{equation}
\end{subequations}
and
\begin{subequations}
\begin{equation}
	\gamma = -j \frac{e^2 k_B T}{\pi^2 \hbar^2} \log\{2[1+\cosh(\frac{\mu_c}{k_B T})] \}\,,
	 \gamma_D = j \frac{v_f}{2\pi \omega \tau}\,,
	 \end{equation}
	 \begin{equation}
	  D_{\sigma} = 1+ \gamma_D \Delta k_t^2\,,\\
	\Delta = \frac{-2\pi}{v_f k_t^2}(1-\frac{\alpha}{\sqrt{\alpha^2 - v_f^2 k_t^2}})\,,
	\end{equation}
	\begin{equation}
	 R(k_x,k_y) = \dfrac{\alpha + v_f k_x}{\sqrt{\alpha^2 - v_f^2 k_t^2}}\,,\alpha = \omega - \frac{j}{\tau}\,,
	 \end{equation}
	 \begin{equation}
	k_t = \sqrt{k_x^2 + k_y^2}\,,\, k_q = \sqrt{k_x^2 - k_y^2}\,,
	\end{equation}
\end{subequations}
where $\omega$ is the angular frequency, $k_B$ is Boltzmann constant, T is temperature, $e$ is electron charge, $\hbar$ is reduced Planck constant, $\tau$ the phenomenological relaxation time, $v_f$ the
Fermi velocity, and $\mu_c$ is graphene’s chemical potential. 
\par The graphene conductivity modulated by y-direction drift current can be modeled by Doppler shift formulation \cite{morgado2018drift}
\begin{subequations}
	\begin{equation}
	\sigma_{xx}^{d} (\omega,v_d,k_x,k_y) =  \frac{\omega}{\omega - k_y v_d}\sigma_{xx}^{BGK}(\omega - k_y v_d,k_x,k_y)\,,\end{equation}
	\begin{equation}
	\sigma_{xy}^{d} (\omega,v_dk_x,k_y) = \frac{\omega}{\omega - k_y v_d}\sigma_{xy}^{BGK}(\omega - k_y v_d,k_x,k_y)\,,\end{equation}
\begin{equation}
	\sigma_{yx}^{d} (\omega,v_dk_x,k_y) = \frac{\omega}{\omega - k_y v_d}\sigma_{yx}^{BGK}(\omega - k_y v_d,k_x,k_y)\,,\end{equation}
	\begin{equation}
	\sigma_{yy}^{d} (\omega,v_dk_x,k_y) = \frac{\omega}{\omega - k_y v_d}\sigma_{yy}^{BGK}(\omega - k_y v_d,k_x,k_y)\,,
\end{equation}
\end{subequations}
where $v_d = 0.0-1.0\, v_f$. The surface conductivity is magnified along y-direction and diminished along the negative-y-direction.
\par
For the densely packed graphene ribbon array which has a unit cell size much smaller than the wavelength, the effective medium theory is applied to calculate the surface conductivity tensor \cite{luukkonen2008simple}
\begin{subequations}
	\begin{equation}
	C_{e} = \epsilon_e \epsilon_0 \frac{2L}{\pi} \log(\frac{1}{\sin(\pi\frac{L-W}{2L})})\,,\, \epsilon_e = \frac{\epsilon_{t1}+\epsilon_{t2}}{2}\,,
	\end{equation}
	\begin{equation}
	\sigma^e_{xx} = (\frac{1}{\sigma^d_{xx}} - \frac{j}{\omega C_e})^{-1}\,,\,
	\sigma^e_{xy} = \frac{W}{L}\sigma^d_{xy}\frac{\sigma^e_{xx}}{\sigma^d_{xx}}\,,\end{equation}
	\begin{equation}
	\sigma^e_{yx} = \frac{W}{L}\sigma^d_{yx}\frac{\sigma^e_{xx}}{\sigma^d_{xx}}\,,\,\sigma^e_{yy} = \frac{W}{L}\sigma^d_{yy} - \frac{W}{L}\sigma^d_{yx}\frac{\sigma^d_{xy}}{\sigma^d_{xx}} + \sigma^e_{xy}\frac{\sigma^e_{yx}}{\sigma^e_{xx}}\,,
	\end{equation}
\end{subequations}
where W is the ribbon's width, L is the length of the unit cell, $\epsilon_{t1}$ and $\epsilon_{t2}$ are the transverse permittivity of the layers above and below the graphene ribbon array.
\section{Hexagonal Boron Nitride Permittivity Tensor}
\label{a4h}
Hexagonal Boron Nitride (hBN) is a uniaxial anisotropic material and supports two phonon-polariton modes in the mid-IR range and arises hyperbolic response. Its dielectric properties can be characterized by the anisotropic permittivity tensor as \cite{hajati2019modal}
\begin{eqnarray}
{\underline{\underline{\bm{\varepsilon}}}}=\begin{bmatrix}
\epsilon_{t} & 0 & 0\\
0 & \epsilon_{t} & 0\\
0 & 0 & \epsilon_{z}
\end{bmatrix}\,,
\end{eqnarray}
\begin{eqnarray}
\epsilon_i(\omega) = \epsilon_{i\infty} (1+\frac{\omega_{LOi}^2-\omega_{TOi}^2}{\omega_{TOi}^2 - \omega^2 - j\omega\Gamma_i})\,,
\end{eqnarray}
where i = t, z. $\epsilon_{t\infty} = 2.95$, $\epsilon_{z\infty} = 4.87$, $\Gamma_{t} = 4\, \textrm{cm}^{-1}$, and $\Gamma_{T\infty}=5\,\text{cm}^{-1}$. In the lower phonon-polariton band, where $\omega_{TO} = 780\,\textrm{cm}^{-1}$ and $\omega_{LO} = 830\textrm{cm}^{-1}$, hBN can propagate type $\RomanNumeralCaps{1}$ phonon mode. While in the upper phonon-polariton band, where $\omega_{TO} = 1370 \text{cm}^{-1}$ and $\omega_{LO} =1610 \text{cm}^{-1}$, hBN can exhibit type $\RomanNumeralCaps{2}$ phonon mode. In these hyperbolic bands, the real part of hBN dielectric function is negative. We choose $f = 7.685THz$ located at the phonon-polariton resonance frequency of the type $\RomanNumeralCaps{2}$ phonon mode, so that the hBN $\epsilon_t \approx 0$. 
\newpage
\bibliography{apssamp}
\end{document}